# Mass Gatherings for Political Expression Had No Discernable Association with the Local Course of the COVID-19 Pandemic in the USA in 2020 and 2021

Eric Feltham[1,2]*, Laura Forastiere[1,3], Marcus Alexander[1,4], Nicholas A. Christakis[1,2,5,6]

**Affiliations**
[1] Yale Institute for Network Science, Yale University; New Haven, CT
[2] Department of Sociology, Yale University; New Haven, CT
[3] Department of Biostatistics, Yale School of Public Health; New Haven, CT
[4] Frank H. Netter MD School of Medicine, Quinnipiac University, North Haven, CT
[5] Department of Statistics and Data Science, Yale University; New Haven, CT
[6] Department of Medicine, Yale School of Medicine; New Haven, CT

*Corresponding author. Email: eric.feltham@yale.edu

**Abstract:** Epidemic disease can spread during mass gatherings. We assessed the impact on the local-area trajectory of the COVID-19 epidemic of a type of mass gathering about which comprehensive data were available. Here, we examined five types of political events in 2020 and 2021: the US primary elections; the US Senate special election in Georgia; the gubernatorial elections in New Jersey and Virginia; Donald Trump's political rallies; and the Black Lives Matter protests. Our study period encompassed over 700 such mass gatherings during multiple phases of the pandemic. We used data from the 48 contiguous states, representing 3,119 counties, and we implemented a novel extension of a recently developed non-parametric, generalized difference-in-difference estimator with a (high-quality) matching procedure for panel data to estimate the average effect of the gatherings on local mortality and other outcomes. There were no statistically significant increases in cases, deaths, or a measure of epidemic transmissibility ($R_t$) in a 40-day period following large-scale political activities. We estimated small and statistically insignificant effects, corresponding to an average difference of -0.0567 deaths (95% CI = -0.319, 0.162), and 8.275 cases (95% CI = -1.383, 20.7), on each day, for counties that held mass gatherings for political expression compared to matched control counties. In sum, there is no statistical evidence of a material increase in local COVID-19 deaths, cases, or transmissibility after mass gatherings for political expression during the first two years of the pandemic in the USA. This may relate to the specific manner in which such activities are typically conducted.

## Introduction

COVID-19, like any serious outbreak of a contagious disease, can place the virtues of public health and civic engagement – in the form of mass gatherings – into direct conflict[1]. Indeed, epidemics pose an especially difficult problem for the political process, which relies on an active and engaged citizenry so that the public may hold their leaders accountable for policies enacted to manage the epidemic in the first place. Even if the will of the majority is aligned with the public interest of most effectively containing, mitigating, and recovering from an epidemic, political participation is often necessary to enact the majority's preferences. Yet, when citizens gather in person, participation itself can contribute to a worsening of the epidemic. Both from a public health perspective and an individual point-of-view, the question arises whether in-person mass gatherings – in particular for political purposes – are safe.

In the Spring of 2020, the United States was in the midst of its presidential primary elections when the COVID-19 pandemic began in earnest. Concerns immediately arose about the potential of in-person voting to lead to a rise in community transmission of the SARS-CoV-2 virus, making elections into potential super-spreader events with the adverse consequence of otherwise-preventable excess deaths in communities across the US. Furthermore, election procedures themselves became a partisan issue under COVID-19[2,3], with some Democrats attempting to signal a commitment to public health, by avoiding the polls, and some Republicans signaling defiance of COVID-19, by showing up in person. COVID-19 sparked a partisan uproar over postponing primaries and mail-in voting, reaching its zenith with the Wisconsin Supreme Court battle[4], where a Republican effort blocked an attempt by Governor Tony Evers, a Democrat, to reschedule the primary election, given his public health concerns. By November, 2020, Dr. Anthony Fauci was stating that in-person voting was likely to be safe, so long as social distancing measures are in place[5].

In addition to the primary elections, which were held in the spring of 2020 in all 50 states, President Donald Trump held 67 political rallies in 15 states across the country during the 2020 election season. The Trump rallies were anticipated to cause a surge in COVID-19, and were among the largest events held in the US at the time[6], drawing between hundreds and thousands of people[7]. The US also saw a surge in large protests for racial justice in the summer of 2020, and sporadically thereafter. The "Black Lives Matter" (BLM) protests (of which there were hundreds) have been characterized as the largest concerted protest movement in US History[8], and cities and towns across the country saw demonstrations ranging in size from tens to hundreds of thousands of people. Early on, some experts worried that the BLM protests were a risk for increasing COVID-19 transmission while the US was in the middle of its second wave of the epidemic[9]. At the same time, over 1,000 public health workers controversially signed a letter that minimized the health risks of participating in the protests, arguing that the political issue of racial justice outweighed any public health risks from COVID-19[10,11].

Political activities as a type of mass gathering during the COVID-19 pandemic continue to be a contentious issue and pressing policy problem around the world.

For instance, in late 2021, a wave of teachers were reported to have tested positive and died subsequent to working at the polls in India[12], and Argentina decided to push back its 2021 midterm elections, while nine countries in Latin America held other elections in 2021[13]. Leaders around the globe continue to decide whether to carry out or postpone their upcoming elections and wrestle with rules for voting[14,15,16,17]. Additionally, large protests continue around the world in places where the pandemic is ongoing, including in India, Columbia, Cuba, and others[18]. Cuba had a record-breaking number of daily COVID-19 deaths around the time of its protests[19]. In India, the government has urged the dissolution of large-scale protests, citing the dangers of COVID-19 transmission[20]. In Russia in February and March, 2022, anti-war protests were banned and COVID-19 was used as the justification[21]. In most of these places, powerful vaccines remain scarce and traditional non-pharmaceutical interventions (NPIs) (including a ban on mass gatherings) are the primary strategy to mitigate the spread of COVID-19.

Political activities are thus an important subset of mass gatherings that have occurred during the COVID-19 pandemic, with some properties that can sometimes make them distinct from concerts, sporting events, or other gatherings marked by more sustained or intimate or indoor contact. Still, the political gatherings studied here are crucial to the functioning of democracy, and are important in their own right from a public health perspective. Moreover, it is possible to comprehensively enumerate certain types of political mass gatherings and to assemble data about them. We thus examine political gatherings both because it is possible to collect complete data on such gatherings and because of their special importance.

## Results

### Methodology Summary

We assembled a complete dataset of essentially all mass gatherings of particular types for a two-year period (see Methods for data availability). We examine the impact of five different sorts of significant political events on the spread of COVID-19 in the US, in the period from March 17, 2020 to November 2, 2021: the US Primary Elections (March 17, 2020 to August 11, 2020); the Georgia Special Election (January 5, 2021); the gubernatorial elections in New Jersey and Virginia (November 2, 2021); Trump's 2020 campaign rallies (June 6, 2020 to November 2, 2020); and the BLM protests (Summer 2020).

We cover every state-wide election over our whole study period in which there was a significant amount of in-person voting (and exclude the national general election on November 3, 2020). We include 100% of the official Trump rallies (67 events). For the BLM protests, we specifically examine 100% of the large-scale protests in the time period considered – a total of 658 county-level events with a crowd of at least 800 persons, of which 94% are directly linked to the BLM movement. The events we study thus span the types of political activities engaged by the public; the geographic area of the US; and the various phases of the epidemic, ranging from early onset in March 2020, the second wave over the summer of 2020, the peak of the third wave in January 2021, and the fourth wave in 2021, in which the Delta variant became the dominant strain in the US.

The preponderance of studies that examine the effectiveness of various NPIs and the impact of in-person gatherings on the spread of the SARS-CoV-2 virus to date only consider the Alpha variant[22,23]. However, the Delta variant was found to be between 40-80% more infectious than the initial strain[24,25], and it caused a marked increase in hospitalizations and deaths, despite the increasing administration of vaccines by the time it emerged[26]. The increased transmissibility of new variants may fundamentally alter the relative effectiveness of various NPIs including the banning of mass gatherings. Therefore, we have included in our analysis events taking place in late 2021 in order to examine the impact of the same type of event under different predominant strains of the virus. Furthermore, our analyses are naturally stratified by vaccine period. In the US, the first vaccines were administered to the public on December 14, 2020[27]. The 2020 Primary Elections, the BLM protests, and the Trump campaign rallies all predate the vaccine rollout in the United States (the last rally in our data was held on November 02, 2020). Similarly, The GA Special Election took place on January 05, 2021, at which time under 1% of the population in the state of Georgia had their first vaccine dose. By contrast, the NJ and VA gubernatorial elections were held when 64 and 62% of those states' populations had complete series (two doses of a two-dose vaccine, or one-dose of a single-dose vaccine), respectively[28].

Modeling the spread of COVID-19 is particularly challenging for two major reasons. The first relates to the quality of the underlying data, which can be incomplete or noisy. The US has struggled to test an adequate number of individuals, and, especially early on, there were not enough tests to adequately track the number of cases. Many of the mass gatherings in our data occurred during a rapid increase in the number of tests administered (see Supplementary Fig. 1), which also occurred unevenly across US counties. Moreover, testing relies on the interest and ability of individuals to get tested and of institutions to test them. Consequently, there is substantial variation in testing capacity and rates at the county level, as well as over time[29]. Furthermore, given that 66% of registered voters reported being at least somewhat concerned about the safety of voting in a March 2020 survey[30] and evidence that individuals may have altered their behavior in response to the BLM protests[31], it is plausible that testing may be endogenous to treatment; that is, individuals in counties that held political events may have been more likely to seek testing, thus potentially inducing bias in the outcomes measured across counties. Consequently, to track the course of the epidemic, we primarily focus on (more accurate) *mortality* data – using both death counts and rates. But we also explore other outcomes, including case rates; virus transmissibility ($R_t$) (from an epidemiological model tailored to COVID-19)[32–35]; and mobility (based on mobile phone data).

Second, many of the available statistical tools make assumptions that are not appropriate to modelling the spread of a virus[36]. Consequently, we developed and applied a non-parametric generalized difference-in-differences estimator for panel data[37], to account for the non-linear epidemic curve, with a matching procedure at the county level (see Methods for methodological details and code availability). We make use of county-level data, as it is the most granular COVID-19 data widely available in the US, affording the best possible means to track changes in the spread of the virus. For each time point (day), we matched “treated” counties (that had a political gathering) with multiple “control” counties (that did not have any political gathering during a sufficiently long time window) that had similar socio-demographic characteristics and similar dynamics of the epidemic before and up to the time of the “treatment.” Crucially, to capture the epidemic characteristics, we matched on the county-level cumulative death rate and also the date of the first reported case in a county. We select, at most, the five best matches to a treated county as the basis for comparison which is a widely used strategy for methods with multiple matches[38]. Calipers were applied individually to specific matching covariates as needed in order to ensure that covariate (im)balance remains below 0.1, on average, over the matching period. Consequently, fewer than five matches are used when a match is of poor quality. Treated observations were dropped from the analysis if they had no good quality matches. Supplementary Fig. 2 displays the distribution of the number of matches for the main analyses; see the Supplementary Information (Section 3) for details on the numbers of matches and treated units for each model presented. Additionally, we present the results for each analysis prior to refining the matches to, at most, the best five, presented in the Supplementary Information, Section 3. Indeed, we conducted a broad variety of robustness checks regarding the quality of this matching – see Methods and the Supplementary Information (Section 2) for further details; we present robustness results with varying caliper widths, pre-outcome-window results, and different matching covariate specifications.

Our approach extends a statistical approach that has been recently introduced and validated[37]. In particular, we define different “causal” parameters, more meaningful for this setting, and we implement a sliding time window to find matches for units that can be considered “treated” at each time point (see Methods). This approach avoids certain limitations in other methods previously applied to assess the effect of mass gatherings on the spread of COVID whose assumptions may not be well met in this setting[39,40]. The present approach may be applied in any setting governed by a non-linear contagion process, whether the “treatment” occurs in a limited time frame or is repeated over time (time-varying treatment) or whether its occurrence varies across the population.

Furthermore, each of the event types we analyze presents different challenges that benefit from alteration of our estimation procedure; in the case of the Trump rallies, for instance, we assume the presence of geographic spillover effects[41], and, in the case of the BLM rallies, we have many units with multiple treatment-like protests within a narrow window of time.

We measure the average effect of an event on the observations where the event actually occurred (the so-called “average treatment effect on the treated” or ATT), by comparing the change in outcome over time across matched units in different treatment and control groups. The difference-in-differences strategy requires that, in the absence of the event, the outcome would have followed the same trend in the treated and control arms. Nonetheless, our approach makes less stringent assumptions than many traditional approaches for the estimation of causal effects with panel data[37]. As opposed to the standard difference-in-differences estimator, this method relies on a parallel trend assumption only after conditioning on both baseline and time-varying covariates before the intervention, including the pre-treatment outcome history.

Covariate adjustment is conducted using a matching procedure whereby the five best matches are selected and used for each treated county (see Methods for details). For each model, the quality of the five best matches was examined with reference to a balance score, calculated based on the standardized distance between a matched unit and its best matches in terms of covariate similarity. In each case, acceptable balance scores were obtained, on average, for each matching covariate (see Methods and the Supplementary Information Section 2 for detailed reporting on match quality for each analysis). This approach also allows for time-varying treatments that can occur multiple times over the observed window.

We used a range of variables describing the counties, including demographic, political, and mobility variables. Our mobility data tracks daily visits to points of interest at individual locations across the US in census block groups, typically covering between 600-3,000 individuals, aggregated to our 3,119 counties (see Methods). Mobility data has previously been linked to the spread of COVID-19[41–43]. Many existing studies use general movement patterns[44–46], such as overall time spent away from home, which does not necessarily indicate whether individuals engage in the sort of close interpersonal contact that is heavily responsible for the spread of COVID-19[42]. However, more recent studies have used finer-grained mobility metrics[42,43,47,48] in line with our approach. We track movement specifically at full-service restaurants, grocers, and fitness and recreation facilities, some of which are known to be high-risk locations[43,47], and directly include these measures in the matching procedure.

In our analysis, a unit is the entire time series of a US county, $X_i$. A unit is treated on a specific day, $X_{i,t}$, if it holds an event (mass gathering) of interest (e.g., an election, rally, or protest) on that day. For such a given treated observation, we match on that county's characteristics for 30 days up to 1 day before treatment. Our primary outcome of interest is the county-level death rate (measured in deaths per 10,000 people, per million people, or in absolute counts of deaths). We contrast the treated and matched unit's outcome for each day in an outcome window defined as 10 to 40 days *after* an event, which is the epidemiologically-informed period in which we would expect an effect on the *mortality rate* from an event where contagion with SARS-CoV-2 might occur[49,50]. This window captures the first-wave of deaths that would be expected to occur from infection on the day of a mass gathering.

For consistency, we us the same time window for the primary analyses of the other outcomes, too (see Methods). That is, we use the same window for cases. While the incubation period is shorter than 10 days, we believe that this window effectively accounts for reporting delays in testing, and the period from 10 to 40 days can capture subsequent waves of infections that could be traced to infections on the day of an event.

For each day that we estimate in the outcome window, we define the "crossover window" as the period in which we examine the treatment history of potential matched control counties. A control county may not be treated during the post-treatment portion of the crossover window, defined as 30 days before the given day in the outcome window (a day *F* between 10 and 40), up to the day before treatment. Eligible control units must have treatment histories that are similar to the treated unit in the pre-treatment crossover window. Thus, we handle the treatment histories in a nuanced manner, separately from the fixed covariate matching window (Fig. 1). Additionally, for each of the main analyses presented, we also calculate the ATTs in the crossover window and up to the start of the outcome window (in these cases, we observe results that are consistent with those in the outcome window itself; see Extended Data Fig. 1).

Valid estimation in this context requires that there is sufficient variation of the treatment onsets to find a suitable set of matches for each treated unit. For example, starting with an analysis of the primaries, the rolling schedule of these primaries in the US, from February to August 2020, helps to satisfy this condition. Furthermore, while states are not randomly assigned to primary election days, and while states with later elections may differ in their characteristics from those with earlier elections, we observe sufficient county-level variation in the features of interest to obtain acceptable matching to the treated units (see Methods and the Supplementary Information Section 2). While several states had primaries that were either cancelled or rescheduled due to COVID-19, it seems justified to treat cancelled primaries as control units over those dates. Furthermore, for those primaries that were rescheduled, the overall difference in the case and death rates, between the original and rescheduled dates, is close to zero (Supplementary Fig. 3). The rescheduled primaries were moved to periods in which community spread still obtained, allowing us to use the counties therein as valid treated units.

### Overall Assessment of Local Mortality and Cases

We first report a summary analysis that examined the overall effect for *all* event types and time periods combined, using county-level death rates and case rates (Fig. 2A), which correspond to an average ATT of -0.257 (95% confidence interval (CI) = -3.482-2.777, p-value (P) = 0.853, Bayes Factor (BF) = 0.030) deaths per million persons for the death rates, and an average ATT of -20.949 (95% CI = -80.987-27.604, P = 0.414, BF = 0.038) cases per million persons. The point estimates on these effects are, in fact, negative. The high quality of the covariate balance between cases and controls for this omnibus analysis is shown in Figs. 2B and 2C.

We emphasize that, in most cases, the changes in the wake of the events are small in magnitude in absolute terms as well. In the context of the death counts, the foregoing rate translates to an ATT for the death *count* of -0.0567 (95% CI = -0.319-0.162); such that 0.0567 more deaths would have occurred, on average, for each day over the outcome window in the counties that held events, compared to a situation in which the political events had not been held.

We stress that mortality is the more reliable and less biased metric (since people might be motivated to modify their testing behavior before or after events), as we found that there was a significant increase in the total number of tests performed over each day of the outcome window and many of the earlier events occurred during a rapid ramp-up in testing at the national level.

**Specific Types of Events and Local Mortality and Cases**

We first analyzed the primary elections of the spring and summer of 2020. Our estimates for the effect of the primary elections, for each day in the outcome window, throughout the US, are generally not significant at the 5% level (Fig. 3A). Furthermore, the averages across the *entire* outcome window for the difference in the change in deaths per million persons (ATT = 0.180; 95% CI = -1.256-1.927, P = 0.827, BF = 0.037) and cases per million persons (ATT = -8.221, 95% CI = -47.524-32.47, P = 0.667, BF = 0.038) were not significant (Supplementary Table 2), in keeping with the day-by-day results.

Because the number of voters present at the polls – in particular, the number of voters in a tightly congested area – might affect the risk of transmission, we did additional analyses stratified on the turnout rate. We observed that no level of turnout yields a positive and significant increase for either outcome compared to the matched control units (Fig. 3B). Separately, we conducted additional analyses that stratified on the population density, primary date, census region, 2016 Trump vote share, the number of days between the first reported case and the primary date, the cumulative case rate on the primary day, and the cumulative death rate on the primary day for the treated counties (Supplementary Figs. 4-17), and we consistently found no clear evidence for an effect of the primary elections on COVID-19 in the subsequent period.

Second, we estimated the effect of the January 5, 2021 US Senate Runoff elections in Georgia. This election is notable both because it (1) occurred during the third peak of COVID-19 in the US, when the rate of new cases was at the highest point during the pandemic; and (2) was extremely competitive and high-profile, attracting attention at the national level[51]. The latter meant that this primary election set a record for voter turnout[52] and the mean in-person turnout rate was higher than that of the primary elections (Supplementary Fig. 18A-B). Yet, despite these differences, and consistent with the primary election results, we observe a mixture of insignificant and positive significant effects on a day-by-day basis (Fig. 4A). But the outcome-window-wide average for the death rate represents an average growth in the death rate on the order of 10 deaths per million persons, and is not significant at the 5% level (ATT = 9.197, 95% CI = -8.972-29.541, P = 0.385, BF = 0.141). The case rate outcome average was also not significant (ATT = -69.949 per million persons; 95% CI = -325.136-175.960, P = 0.552, BF = 0.105) (Supplementary Table 2).

Furthermore, models that stratified by in-person voter turnout do not reveal significant differences in the ATTs as the turnout rate increases (Supplementary Figs. 19-20). Moreover, we conducted an additional analysis that stratified by the county-level propensity to wear masks; this analysis did not yield significant effects, on average, across the outcome window, although the effect size diminished by half in counties that had a tendency to report mask-wearing above the median for counties in Georgia (Supplementary Figs. 21-22; Supplementary Table 2). Finally, a similar effect was observed in an analysis that stratified by above or below 50% vote share for Donald Trump in the 2016 election (Supplementary Figs. 23-24).

Third, we assessed the effect of the November 2, 2021 gubernatorial elections in New Jersey and Virginia. These elections had in-person turnout rates that were over twice that of the primary elections (Supplementary Fig. 18C-D), and, like the election in Georgia, were prominent on the national political stage and set records for levels of turnout[53,54]. Additionally, these elections were held during the surge of the Delta variant in the United States, and occurred one month prior to a new set of policy responses to mitigate the spread of the Omicron variant. The election was also held during a period of concern about "pandemic fatigue" (drops in adherence to social distancing and masking protocols as frustration with pandemic restriction forms in the public was rising)[55–57]. While we could not directly match on the county-level vaccination status in these late elections, our included demographic characteristics are strongly associated with the vaccination levels in the US[58,59]. Here again we observe no significant evidence for an effect of the elections on the mortality rate or case rate after the election in both states combined, with an outcome-window-wide average ATT for the death rate of -15.927 per million persons (95% CI = -29.19- -4.298, P = 0.013, BF = 2.472) and insignificant average for the case rate (ATT = -220.749 per million persons, 95% CI = -482.639-135.296, P = 0.139, BF = 0.296) (Fig. 4B; Supplementary Table 2).

Fourth, we estimated the effect of Donald Trump's political rallies, held over the course of 2020. These rallies varied in attendance, and they were held outdoors in all but three cases (Tulsa, June 20; Phoenix, June 23; Henderson, September 13). The rallies themselves were most commonly held at airports. Estimates for the rally sizes range from hundreds to thousands[7], although no precise or consistent data are available. We assumed that individuals were willing to attend a rally from up to three counties away from the rally location, and consequently apply a model that accounts for spillover effects onto neighboring counties. Specifically, we define four levels of exposure to the treatment: (1) direct treatment, where a county holds a rally, (2) 1$^{st}$ degree exposure, where a county borders one that holds a rally, (3) 2$^{nd}$ degree, where a county is one county away from one that holds a rally, and (4) 3$^{rd}$ degree, where a county is two counties away from one that holds a rally. In each case, we find no positive and significant set of estimates over the outcome window, 10 to 40 days after a rally, for each level of exposure to a rally, with insignificant outcome-window-wide averages for the death rate (ATT = -0.333 per million persons, 95% CI = -6.157-5.07, P = 0.983, BF = 0.144) and case rate (ATT = 10.636 per million persons, 95% CI = -169.233-213.967, P = 0.984, BF = 0.137) for direct treatment (Fig. 5A; Supplementary Table 2). We also examined indirect exposures in neighboring counties and found no evidence of an association for death and case rates (Fig. 5B). Additionally, we stratified by Trump's share of the vote; we supposed that counties that had a majority vote for Trump would be more likely to contain rally attendees, and possess a greater likelihood of rally-induced transmission. In each stratum, we found no positive and significant pattern over the outcome window (See Supplementary Figs. 25-28). Further, we note that it is plausible that individuals were infected but not did not seek testing after attending

Trump rallies, which further underscores the death rate as a more reliable metric of interest in this context.

Fifth, we estimated the effect of the BLM protests, and found no evidence for an effect of the protests on the mortality rate, with insignificant outcome-window-wise averages for death rate (ATT = -0.442 per million persons, 95% CI = -2.833-1.836, P = 0.688, BF = 0.058) and case rate (ATT = 31.380 per million persons, 95% CI = -12.475-85.095, P = 0.116, BF = 0.018) (Fig. 6A; Supplementary Table 2). Additionally, we stratified by the total crowd size estimate for all protests on a given day in a county, and we do not find evidence of an effect of the protests across the strata, even for the largest protests (ranging from around 3,000 to 30,000 persons) (Fig. 6B). We also found no clear and significant effects in a model that stratified treated observations based on the number of recent protests (that totaled above 800 persons) that occurred within the three weeks prior to the focal protest (see Supplementary Figs. 29-30). See the Supplementary Information Sections 1 and 2 for more details and for further robustness checks). Extended Data Figs. 2-9 present the estimates prior to refinement and with and without calipers for the death rate models.

### Political Gatherings and Local COVID-19 Transmission ($R_t$)

In addition to analyzing the effect of these five event types on the case and mortality rates, as discussed above, we also applied our approach to estimates from an independently developed epidemiological model of the transmissibility of SARS-CoV-2 which reconstructed the course of the virus from the observed deaths and cases, and from epidemiologically-informed assumptions about the dynamics of the virus (see Methods for further details)[60]. We examine the impact of each of the five event types on the county-level estimated effective reproduction number ($R_t$). In each situation, we observe no statistically significant impact of large-scale political events on the difference in the change in $R_t$ from the day before an event to between 0 and 20 days afterward between treated and matched counties (Fig. 7). Note especially the lack of estimated impact on the day of an event, or in the immediate period within ten days of the event. Furthermore, the effects are close to zero in magnitude in the initial few days, with small confidence intervals, which further increases our confidence in the absence of an effect on the transmissibility of SARS-CoV-2 from the political events we have studied.

### Political Gatherings and Local Mobility

Finally, we applied our method to examine the impact of the political gatherings on our three mobility metrics (at fitness and recreation centers, full-service restaurants, and grocers) known to be associated with the spread of SARS-CoV-2[43,47]. In each case, we find no significant impact of these political events on visits to these location types over an outcome window of 0 to 20 days after an event. That is, there are no statistically significant changes in overall visits to these location types on the day of the events themselves, nor a discernable impact on mobility patterns subsequent to the events (Fig. 8).

### Discussion

Across a broad array of mass gatherings – including elections, rallies, and protests that took place throughout 2020 and 2021 – we find no substantial evidence for a material deflection of the local course of the pandemic, for a period from 10 to 40 days afterwards. Our analysis here is *ecological*[61], in the sense that we examine whether political activities affected the course of the pandemic in a county, not whether, at the *individual* level, going to the polls or participating in a rally or protest affects an individual's risk of contracting the infection.

There is no fixed, intrinsic connection between mass gatherings of a specifically political nature and the spread of SARS-CoV-2 per se. Our focus on political gatherings was driven partly by their particular importance and partly by the fact that it was possible to systematically assemble a comprehensive database of all such gatherings, with precise timings (in a way that would not have been as feasible had we focused on other sorts of gatherings, like sporting events, musical concerts, and so on).

Furthermore, the impact of the gatherings on the epidemic – especially in the case of elections – hangs on the particular policies and behaviors found at individual voting sites, for instance, whether masks are used by poll-workers and voters, whether most of the time is spent outside, and whether individuals are physically distanced or quiet while waiting in line. Voting is indeed not ordinarily an intimate activity. And the prevalence of the virus in a given time and place also likely plays a role. Super-spreader events do happen, especially with interpersonal contact[62]. For instance, a single wedding in Maine in August, 2020 was linked to 270 cases and eight deaths[63], enough to fundamentally alter the trajectory of COVID-19 in a state that previously had relatively few deaths and cases. However, many of these super-spreading events during the COVID-19 pandemic have been in familial settings associated with close contact and a lack of precautions, or have involved sustained indoor exposure[62,64]. Hence, voting or other political activities may present a similar level of risk to shopping at a grocery store or waiting in line outside a restaurant for a take-out order. In fact, on many days during the outcome window for the NJ and VA gubernatorial elections, we even observe significant and *negative* estimates, which may reflect the fact that voting could actually be a safer alternative to usual activity patterns then prevalent (which might have been more likely to involve direct, close interpersonal contact).

While we observe a lack of consistent evidence for a spike in mortality associated with the primaries, Georgia special election, or the New Jersey and Virginia gubernatorial elections, it is important to note several caveats with respect to the safety of voting in general. Most of the primary elections were held under fair-weather conditions, enabling individuals to wait in long-lines outside, at lower risk of transmission – although, this was less true for the Georgia Special Election (that took place in January) or the New Jersey and Virginia elections (in November). It remains possible that the effect is moderated by the tendency to wear masks or socially distance at the polls. We expect that mass gatherings in close conditions without social distancing, mask use, or vaccination with boosters may pose a more significant risk of transmission, especially in cases of high turnout where voting takes place in crammed and enclosed spaces.

Troublingly, while policy arguments about mass gatherings have been politicized during the COVID-19 pandemic, the discussion has largely taken place in the absence of a comprehensive assessment of the public health risks of collective political activities. With respect to the US elections, three studies have been conducted on the Wisconsin primary election; and they reached conflicting conclusions regarding the impact of in-person voting on the subsequent course of the COVID-19 epidemic thought two found no subsequent increase[65–67]. Separately, two studies have been conducted on the effect of the BLM protests, each finding no clear effect[44,68]. Our comprehensive study of the whole US and a two-year period thus sheds light on this topic.

Smaller-scale studies in foreign countries also paint a mixed picture, though most comport with our findings. Examining the first and second rounds of the 2020 French municipal elections, one study found a sizeable subsequent increase in hospitalization rates after the first-round election (in March) but did not find any increase associated with the second-round elections (in June) when masks became obligatory[69]. Two further studies using, respectively, a Bayesian mixture model and a sigmoidal mixed effects model found no effect of the same French elections for either round[70,71]. A study of the October, 2020 elections in the Czech Republic found a subsequent increase in cases[72]. Still another study of elections in Italy in September, 2020 found that a subsequent increase in the spread of COVID-19 was driven mainly by campaign activities leading up to the elections, which allowed indoor events at public and private places with no limit on the size of the event (in contrast to size restrictions imposed on other event types in Italy at the time), rather than by infections at the polls themselves[73]. In still another context, researchers found no effect of voting on the spread of COVID-19 in the 2020 Brazilian Municipal Elections, which occurred in the context of broad public health campaign and restrictions including a mandate that voters wear masks at the polls[74]. Finally, an additional study found no increased mortality among 163,000 candidates in the 2020 French Town Hall elections[75].

Generally, the studies that have found significant effects of political events are those that either do not account for the non-linear contagion process[66,76], or do so only indirectly[69,72]. By contrast, studies that directly account for the contagion process, like the present one, have found no significant effects[65,70,71]. Our study thus adds to the literature by directly accounting for the non-linear contagion process in a generalized differences-in-differences framework, and avoiding some key difficulties associated with some epidemiological models[77] and many existing causal inference methods[36].

In contrast to voting, the Trump rallies and BLM protests were almost entirely held outdoors in the open-air, despite large crowd sizes. Outdoor transmission of COVID-19 is very rare[78]. The Sturgis Motorcycle Rally is often cited as an example of outdoor transmission[79]; however, it was marked by indoor events in addition to the outdoor gathering, with cases stemming from the rally linked to restaurants and workplaces[80]. Consistent with this finding, we found no evidence of an uptick in visits to restaurants or other locations known to yield a high risk of transmission associated with the protests or rallies we studied (Fig. 8; Supplementary Figs. 31-32). Furthermore, it is also possible that the average age of people attending rallies and protests is younger, and their underlying health better, than the electorate as a whole.

Another possible explanation of the observed results with respect to the mass gatherings we have studied is that people in an area with recent political activity, or the individuals participating in such activity themselves, may make compensatory adjustments to lower their *other* risks for contracting the virus. For instance, elderly people in an area of a rally or protest may decide to stay inside and avoid contact with others for a few weeks after the rally (even if they themselves did not participate) [81] . On this account, the fact that the mass gatherings studied here did not seem to deflect the overall COVID-19 mortality curves in the short run could be driven by compensation for the risks of the events themselves[44]. However, we found no evidence for mobility decreases following the large-scale political activities (at locations known to be at high risk for transmission) that would support such an account (Fig. 8; Supplementary Figs. 33-34).

Our study is potentially limited by unobserved factors, as with any statistical approach to observational data, including prior analyses of COVID-19 risks and responses[40,43,44,47,50,65,66,77,79]. Such inferential problems may indeed be worse for a process that is intrinsically contagious, which is one of the reasons we also sought to match index counties with multiple control counties that had similar prior epidemic trajectories. Furthermore, while the results are consistent across time and event types, the confidence intervals encompass a range of possibilities. Finally, these results, focused on a particular class of mass gatherings, are at odds with some epidemiological studies of NPIs (which have included mass gathering bans as a possible stratagem)[50,82–84]. On the other hand, our results are in keeping with two prior studies of some of the BLM protests[44,68], and two prior studies of the Wisconsin primary (i.e., a single state), as noted earlier[65,67]. Furthermore, they are in line with a more complex picture that is being established as researchers grapple with the set of existing conclusions drawn from traditional statistical methods[36].

We urge careful interpretation of these results, which may be specific to these events, in the US context. The full-range of potentially confounding factors should be considered in the context of mass political gatherings across the globe, where leaders are deciding whether to hold in-person elections or restrict the right to protest—especially in places characterized by a scarcity of vaccines.

Furthermore, we emphasize the heterogeneous nature of these events themselves, which are associated with differential risk exposure for different groups; for instance, older voters, who are at high risk for COVID-19, tend to vote at the highest rates[85], while younger individuals were more likely to attend BLM protests. Such effects may also potentially bias some of our outcome metrics; for instance, those who attend Trump rallies may be less likely to seek testing, and younger people who attend BLM protests may be less likely to die from contracting COVID-19. Hence, while the mortality data is generally taken to be a better reflection of the underlying state of the COVID-19 epidemic[86], we performed a comprehensive analysis that considered case rates and transmissibility as supplements to the mortality rate. And we observe no evidence for an effect on any of the outcomes.

Another limitation pertains to the potential relationship between the epidemic and turnout at various of the events we study; it is possible that individuals decided not to turn out to these events for fear of exposure. Our approach helps avoid potential bias due to this potential relationship by matching counties based on characteristics of the epidemic (the death or case rate or transmissibility in the period leading up to the event along with the date of the first reported case). Nonetheless, because we did not have access to comprehensive data on the list of attendees for all events, and were not able to adjust the death counts based on exact attendance, and instead adjust based on the county population. Additionally, we do not include extensive data on health system capacities and constraints; however, given structural inequities in our society, these are likely to be strongly correlated with the extensive set of matching covariates we do include[87,88], and hence are likely similar between our treated counties and their matched controls; as a result, this is less likely to bias our results; still, this lack represents another limitation of our study. While low attendance in counties with higher rates of COVID might have reduced transmission, we see this a possible effect modifier, which does not bias the results. Furthermore, we note that these events were large in scale (Extended Data Fig. 10) often involving thousands of individuals at key phases of the pandemic; our analyses that look directly at variation in election turnout and protest size do not yield significant effects.

Policy responses to voting and political expression should be mindful of their potentially disparate impact on otherwise already disenfranchised groups if policy-makers make a potentially spurious effort to reduce risks of COVID-19[89–92]. And we also note that the decisions individuals make and should make for themselves about whether to attend a political gathering are distinct from the overall collective impact. Still, we consistently find no evidence for a material increase in COVID-19 mortality or other epidemic parameters across a wide-range of settings in the US that arise from mass gatherings for political expression. This lack of evidence stands in contrast to the sizeable effects estimated for some other non-pharmaceutical interventions; as a result, our study may help to underscore the importance of a policy focus on NPIs with demonstrated effects[93]. The complex manifestations of this protean respiratory pathogen should not be underestimated.

# Methods

## 1 Data

### 1.1 Johns Hopkins COVID-19 mortality (and case) data

County-level COVID-19 deaths were collected from a repository maintained by the Johns Hopkins Coronavirus Resource Center[86,94], compiled from local and state health agencies. This dataset is a standard in the literature, and is considered the authoritative source of COVID-19 mortality and case data[95]. Our main analyses used the survey of COVID-19 death counts, while supplementary analyses employ positive test results as an alternative outcome for each estimated model. Our main analyses focus on the COVID-19 death counts, rather than positive test results, as our outcome measure due to: (a) limited and changing testing capacity; (b) bias with respect to COVID-19 testing implementation; and (c) the lack of available test data at the county level for many areas of the country. With respect to (b), individuals were only likely to be tested if they showed symptoms, or had the means to get tested; consequently, the tests cannot be viewed as a representative sample of the population.

In our analysis, we assumed that entries prior to the first reported data were zero counts. And we removed observations not attributable to any specific US county. Additionally, we retained observations where the cumulative death counts declined from one day to the next; decreases may occur due to reporting error, such as a correction of an overcount. In these cases, we marked the change from one day to the next as a zero.

### 1.2 COVID-19 transmissibility estimates

We draw on estimates of the effective reproduction number of the virus, over time, at the county level, from a COVID-19 epidemiological modeling project. This model uses a Bayesian framework that accounts for reporting delays, and variability in case ascertainment to reconstruct the course of the epidemic from observed death and case counts[32,33,35]. Model estimates are freely available online from the project website, or reproducible with "covidestim", an R package[96]. We draw the daily county mean estimates from the model as an additional outcome to supplement our analyses of case and mortality rates.

The model (as implemented by Chitwood et al.) was fit using the data on new cases and deaths to model the unobserved transmission rate, from the Johns Hopkins data (described above). The model uses four compartments: a- or pre-symptomatic; symptomatic and not severe; symptomatic and severe; and death. Uninfected individuals become asymptomatic and either recover or move to the symptomatic state, from which they either recover, or move to the severe symptomatic state. Individuals in the severe state either recover or die. Individuals are assumed to only be infected once. The $R_t$ trend is modelled as a log transformed cubic b-spline with knots every 10 days, and with penalties on the first- and second-differences on the spline parameters to ensure a parsimonious fit. The lags between infection to symptoms, symptomatic cases to severe ones, and severe infections, and reporting delays are taken to be fixed, and according to a Gamma distribution, consistent with the literature[50]. Model results were found to be robust under a range of assumptions[32] Case ascertainment varied significantly over course of the epidemic; to account for this, the fraction diagnosed in the model was allowed to vary over time. Further details, including a full list of model parameters, are available in Chitwood et al[32].

### 1.3 SafeGraph

We applied data from SafeGraph Inc., which collects and aggregates data from a range of mobile phone applications to track population mobility. SafeGraph data, including point-of-interest location data, have been used extensively to study population mobility during COVID-19[43,97,98], and found to be demographically representative.[99] We used the Places-Patterns dataset, which tracks hourly visits to a set of Core Places which are points of interest around the US, that contain addresses and North American Industry Classification System (NAICS) codes. We included three business categories previously found to be associated with possible transmission risk: grocers, gyms and fitness clubs, and full-service restaurants[43,47], We aggregated the place data to the daily and county level for each business category. Each mobility variable represents the total daily visits to locations, classified into the above categories, by users in the SafeGraph database, weighted by the ratio of the number of SafeGraph users resident in that county to the total county population, in order to adjust for the sample size in each county[100]. Additionally, we removed all parent points of interest from the dataset, to ensure that no location is counted twice in the dataset (e.g., counting a mall and the stores therein).

### 1.4 BLM protests

The US Crisis Monitor[101] is a curated dataset produced in collaboration with the Armed Conflict Location & Event Data Project and The Princeton University Bridging Divides Initiative, that contains comprehensive data on political protests, violence, demonstrations, and related events in the US for 2020. We systematically consider all protest-related events that occur in the US between May 26, 2020 and October 10, 2020. We consider all protest-related events that occur in the US between May 26, 2020 and October 10, 2020. This period captures the series of protests that initially started during the aftermath of the murder of George Floyd, and extended across the country[101]. We aggregate the individual protest events to the county-day level. Our data includes all significant events that occur in this period, including some protest events not related to BLM. We observe 658 distinct county-level events, where 94% were classified as part of the BLM movement, 7% as "Blue Lives Matter" pro-police, pro-Donald Trump, and 3% as including direct references to COVID-19. The remaining protests involved various other political causes.

The dataset includes accounts from newspaper and traditional media sources on the location, date, and size of protest events. Protest crowd size estimates range from specific numbers (e.g., 300), numeric ranges (e.g., 300-500), to descriptive estimates (e.g., hundreds, thousands, "almost a dozen"). For general range descriptions, we took the geometric mean over the possible range of values it could encompass (e.g., "hundreds" becomes $\sqrt{100 * 1000}$). These approximations serve our goal to separate protests by orders of magnitude. Nonetheless, crowd size estimates from newspaper sources have been found to produce generally accurate estimates of crowd sizes[102]. In our data, multiple protests frequently occur within a county on a single day; in this case, we summed the estimated protest counts over each individual event. Rather than directly rely on protest size estimates, our main analysis applied an event-size threshold to define a binary treatment variable. We only considered a county 'treated' on a given day if the aggregated protest count was above 800 persons. On the other hand, only counties with protests below 400 persons were eligible to be matched control units to a treated county. The distribution of protest sizes was highly skewed (Supplementary Fig. 35).

### 1.5 US elections

Our elections data consist of the rolling schedule of primary dates. During the 2020 Primary Election cycle, several states rescheduled (Table S1) or cancelled their in-person primary elections. Our primary treatment used the rescheduled election dates, and only elections that were held in-person were considered as treated observations. Counties in which mail-in-only primaries were held were eligible to serve as matched control units. We analyzed elections in 1,173 counties drawn from 21 states where applicable elections were held across the mainland US; we drew on the full set of 3,119 counties in the continental 48 states in the mainland US as our source for match units. The 2020 Primary Election cycle ran from February 3 (IA) to August 11 (CT). In our main analysis, we only considered primaries that occurred after March 14; we believe that any primary election held prior to March 1 is not a valid treatment, given the low likelihood of COVID-19 community spread before March, in most locations.

We further estimated a model that stratifies by voter turnout. We personally compiled in-person voter turnout directly from state and county election agencies, where available. Turnout data were not available for CO, IA, KY, MA, MI, MN, MS, MO, MT, NV, NY, OK, PA, RI, SD, UT, WV; consequently, these states were estimated as a separate category in models that apply turnout data (those results are below, but not in the main manuscript).

Additionally, we drew on elections data for the January 5, 2021 US Senate Special Election in Georgia from the US Elections Project[103], including the in-person turnout in each county, the overall turnout rate, and the number of mail-in ballots. Turnout data for the NJ gubernatorial election was drawn from the Department of State of New Jersey's Department of Elections[104], and turnout data for the VA gubernatorial election was pulled from the Virginia Department of Elections[105], In each case, the in-person voter turnout rate is calculated as the number of in-person ballots case divided by the county population.

### 1.6 Donald Trump's campaign rallies

We used a list of Trump rallies compiled directly from traditional media sources, from June 6, 2020 to November 2, 2020, for a total of 67 individual rallies across the country. On some days, multiple rallies were held in multiple locations, and some locations were repeated across the data over time. In each case, we handle a county as treated on a particular day if a rally occurs. Because it is likely that many rally attendees travel across county lines, from rural more Republican counties to the comparatively urban and central rally locations, we also employ US county adjacency data from the US Census[106] to account for spillover effects to neighboring counties.

### 1.7 US Census

We draw county-level demographic data from the 2014-2018 Five Year American Community Survey (ACS) from the US Census Bureau[107]. Specifically, we construct estimates of the percentage African American, percentage Hispanic, log median income, population density (persons per square mile), percentage at least 65 years of age, and Donald Trump's share of the vote in in the 2016 General Election.

### 1.8 *New York Times* masking behavior survey

The county-level tendency to wear a mask was drawn from the a survey conducted by The New York Times[108] conducted between July 2 and July 14, 2020 consisting of over 250,000 individual responses, weighted by age and gender. Respondents were asked "How often do you wear a mask in public and when you expect to be within six feet of another person?", and given the option to answer "never", "rarely", "sometimes", "frequently", or "always". We constructed the tendency to never or rarely wear a mask as the sum of "never" and "rarely" percentages by county. We did not make use of this data for covariate matching in our general analyses since the survey was conducted amid the primary elections. Specifically, we use this data for a supplementary analysis of the GA special election.

### 1.9 Software and code

Analysis was conducted in the Julia programming language[109] (version 1.7.1) with the "TSCSMethods" Julia package (which we are pleased to release), additional paper replication materials are available on GitHub as the "COVIDPoliticalEvents" Julia package. Data cleaning, preparation, and additional analysis was carried out with the R programming language[110] (version 4.0.3). The map in Figure 1 was made with the "urbnmapr" R package[111].

## 2 Generalized difference-in-differences estimation for COVID-19

Modelers of COVID-19 and interventions that affect its spread face a methodological dilemma: whether to apply an epidemiological or causal statistical method. The first includes compartmental models, e.g., SEIR models, which face serious difficulties with identification[77] and reliance upon estimated epidemiological parameters that may change as the pandemic unfolds, which is increasing in the complexity of the process model. In such models, the estimated parameters may identify descriptive quantities, but they may not identify the causal parameters that would undergird an effective policy response.

The second brushes against the limitations of many common statistical approaches. Existing statistical analyses of the impact of NPIs on the spread of COVID-19 apply either interrupted time series design[40] or difference-in-differences estimation[76]. Both of these approaches rely on the assumption that, without the intervention, the spread of COVID-19 could have been predicted from the model fit to the process from the pre-intervention period, ruling out other time-varying factors that could confound the relationship between the interventions and the spread. Furthermore, these approaches assume that the outcome is a linear process, which is often inappropriate in the context of an infectious disease that is characterized by a non-linear trend. In practical terms, these methods may falsely attribute natural changes in the course of the epidemic to a particular intervention.

Specifically, interrupted time series methods typically (1) do not control for an underlying trend not caused by the intervention or rely on specification of its correct functional form[112], such as the underlying contagion process of an infectious disease, and (2) simply rule out the possibility of unmeasured factors causing a given effect. Specifically, Fowler et al. (2021)[39] and Hsiang et al. (2020)[40] attempted to resolve (1) through fixed-effects models that indirectly account for the contagion process as reduced-form models justified on the assumption that the proportion of susceptible individuals approaches unity. However, we do not find this assumption justified in our setting given the absence of reliable testing data in the US (which severely underestimates the number of active cases)[32], and a study time-horizon that ranges from March 2020 to November 2021, over every phase of the pandemic.

On the other hand, difference-in-differences method does control for an underlying trend not caused by the intervention by relying on the common trend assumption.[113] In addition, more advanced difference-in-differences strategies rely on regression adjustment or matching to allow for unobserved time-invariant factors.[114–116] However, traditionally, they do not account for time-varying imbalances, nor the existence of a non-linear data generating process. The approach here is more in line with recent approaches for difference-in-differences methods in less restrictive setting[114,117]. Instead, both difference-in-differences and interrupted time series applied to COVID-19 often assume that cases are log-linear[39,40,44,66,76,79,118], which induces bias, especially when the mean counts are low and overdispersed, which holds true in our county-level daily death data. Our data are also strongly zero-inflated, limiting the utility of traditional count models. Additionally, the fact that many counties have zero deaths over portions of our study time horizon precludes the estimation of traditional epidemiological models to assess the impact of our interventions. Practically, these methods induce bias that may lead to lead to a spurious assignment of causal pathways, which might, in turn, prompt inappropriate policy responses.

To avoid relying on parametric models for the dynamics of COVID-19, in the wake of these difficulties, we modify and apply a non-parametric, generalized difference-in-differences estimator with a matching procedure for time-series cross-sectional data to estimate the average effect of the treatment on the treated (ATT). We do so by extending an approach from Imai et al. (2021)[37] in a novel way. In particular, we define different causal parameters, more meaningful for this setting, and implement a sliding time window to find matches for units that can be considered "treated" at each time point. Our approach may be applied in any setting governed by a non-linear contagion process, where the "treatment" occurs in a limited time frame or is repeated over time (time-varying treatment) and its occurrence varies across the population. Furthermore, each of the event types we analyze presents different challenges that require alteration of our estimation procedure: in the case of the Trump rallies, we assume the presence of spillover effects, and in the case of the BLM rallies we have many units with multiple treatment-like protests within a narrow window of time.

Our difference-in-differences approach measures the average effect of an event on the observations where the event actually occurred (ATT), by comparing the outcome change over time across matched units in different treatment groups. The difference-in-differences strategy requires that, in the absence of the event, the outcome would have followed the same trend in the treated and control arms. Nonetheless, our approach makes less stringent assumptions than many traditional approaches for the estimation of causal effects with panel data. As opposed to the standard difference-in-differences estimator, this method relies on a parallel trend assumption only after conditioning on both baseline and time-varying covariates before the intervention, including the pre-treatment outcome history. In addition, covariate-adjustment is conducted using a matching procedure. This approach also allows for time-varying treatments that can occur multiple times over the observed window. We believe that our approach is novel in applying such a method to epidemic data, and constitutes a new general strategy for research on the effects of single or intermittent events on a contagion.

In our analysis, a unit is the entire time series of a US county (or county equivalent). A unit is treated on a specific day if it holds an event of interest (e.g., an election, rally, or protest) on that day. For such a given treated observation, we match on that county's characteristics for 30 days prior to treatment, up to 1 day before treatment. We contrast the treated and matched unit's outcome for each day in an outcome window defined as 10 to 40 days after an event, which is the epidemiologically-informed period in which we would expect an effect on the mortality rate from an event where contagion may occur. Given the underlying contagion process, it is not reasonable to assume that the effect of any intervention on the epidemic reverses itself, but rather alters the epidemic trajectory in perpetuity. However, we allow units with prior treatments to serve as eligible controls for a treatment event, under specific conditions. For each day that we estimate on the period from 10 to 40 days, matched counties must have a similar treatment history to the treated unit during the pre-treatment crossover window, and have no treatment during the post-treatment crossover window (detailed below). Consequently, we separately match units to a treated observation for each day in the outcome window.

We do not match on treatment history prior to the pre-treatment cross-over period, since we consider deaths after day 40 to be endogenous to the contagion process. The outcome window period represents the first-wave of deaths that would be expected to occur due first-hand infection at the event. Since these first-wave deaths are caused by an exogenous shock—e.g., an election, rally, or protest—they cannot be predicted from the observed death rate alone. However, deaths that occur beyond 40 days are likely second-wave deaths, which are predictable from the observed death rate. Thus, we consider deaths past 40 days to be endogenous to the contagion process, allowing similarity matching based on the death rate to account for the effects of treatment events that occur prior to the pre-treatment crossover period.

Additionally, we restrict our outcome window to the expected first-wave of deaths because estimation over a longer outcome window risks confounding by events that may occur and thereby separately alter the epidemic trajectory in our treated or control counties. Consequently, there arises a risk of misattributing the effect of such events to the treatment considered. We thus face a tradeoff between capturing the long-run effects of an intervention (i.e., an election, rally, or protest) and applying a shorter outcome window to avoid post-treatment confounding. Hence, we restrict our procedure to the period in which 90% of the first-wave deaths would occur due to transmission on the day of treatment.

Thus, we take the treatments to influence the outcome from 10 to 40 days after a given treatment, corresponding to the 5th and 95th percentiles of the distribution of times from infection to death. The distribution of times from infection until death comes from empirical estimates in the epidemiological literature, and is modeled as the sum of two gamma distributions (Fig. 1):

$$(1)\ \pi \sim \text{Gamma}(5.1, 0.86) + \text{Gamma}(17.8, 0.45)$$

## 3 Identification of the average treatment effect on the treated

Define $X_{i,t}$ to be the treatment for unit *i on day t*, valued $X_{i,t} = 1$ when the unit is treated on day *t* (it holds an event of interest), and 0 otherwise. Based on this outcome window, we assume that the potential outcome of each unit *i* at day $\tau$ depends on the event history in a carry-over window, that is, $F_{max}$ days to $F_{min}$ days before $\tau$, with $F_{max} > F_{min}$. That is, the outcome at $\tau$ depends on interventions $X_{i,q}$, with $q \in \{\tau - F_{max}, \dots 0, \dots, \tau - F_{min}\}$ i.e., $Y_{i,\tau}\left(\{X_{i,q}\}_{q=\tau-F_{max}}^{\tau-F_{min}}\right)$.

We then contrast the average potential outcome at time *t+F*, for a treatment at time *t*, with the average potential outcome at time *t+F* without treatment at time *t*, where $F \in \{F_{\min}, F_{\min}+1, F_{\min}+2, \dots, F_{\max}\}$. In particular, our interest lies in the following causal quantity, $\delta(F; F_{\min}, F_{\max})$, the average treatment effect on the treated (ATT):

$$(2)\ \delta(F; F_{\min}, F_{\max}) = E\left[Y_{i,t+F}\left(X_{i,t} = 1, \{X_{i,t-l_1}\}_{l_1=1}^{F_{\max}-F}, \{X_{i,t+l_2}\}_{l_2=1}^{F-F_{\min}}\right) - Y_{i,t+F}\left(X_{i,t} = 0, \{X_{i,t-l_1}\}_{l_1=1}^{F_{\max}-F}, \{X_{i,t+l_2}\}_{l_2=1}^{F-F_{\min}} = 0\right) \mid X_{i,t} = 1\right]$$

with $F \in \{F_{\min}, F_{\min}+1, F_{\min}+2, \dots, F_{\max}\}$. The parameter $\delta(F; F_{min}, F_{max})$ is the causal effect of a treatment event on day *t* on the death rate *F* days after *t*, compared to not having a treatment event on day *t* and in the post-treatment carry-over window affecting day *t+F*, when the sequence of treatment events are the same in the pre-treatment carry-over window, marginalized over the distribution of the treatment history after the treatment event at day *t* conditional on having the treatment at time *t*.

We rely on two key assumptions:

1. Non-interference between units. We assume that there is no interference between counties over the study period. We assume that an event that takes place in a single county does not affect the death rate in other counties, at least over the time frame we consider.

2. Parallel trends, conditional on treatment, covariate, and outcome histories for $F_{max}$ days before the day of the elections. This assumption relaxes sequential ignorability, the condition that the treatment assignment is unconfounded, conditional on the covariate and outcome history up to $t - F_{max}$. Instead, we allow for the presence of unobserved confounding variables. The parallel trend assumption holds conditionally on the treatment, covariate, and outcome histories for $F_{max}$ days before treatment administration.

The ATT is estimated over a range of days $F \in \{F_{\min}, F_{\min} + 1, F_{\min} + 2, \ldots, F_{\max}\}$, corresponding to the outcome (carry-over) window over which we expect treatment effects to occur. We chose $F_{\min} = 10$ and $F_{\max} = 40$, respectively, as to the 5th and 95th percentiles of the distribution of the number of days from infection to death. This window captures the first-wave of deaths expected from an exogenous shock (i.e., an election, rally, or protest).

## 4 County matching procedure

The matching procedure contains three steps: restriction on the treatment history of each county to determine the set of allowable matches, distance matching based on chosen covariates, and matching refinement from the set of allowable matches.

### 4.1 Selection of eligible matches based on treatment history

The definition of our ATT implies that, for treatment day $t$, and outcome day $t$+$F$, treated units (that had the treatment on that day) should be compared with control units that had no treatment events on day $t$ nor during the post-treatment cross-over window for $t$+$F$, i.e., ($t$, $t + F - F_{\min}$). During the pre-treatment crossover window ($t + F - F_{\max}$, $t - 1$), units are matched on their treatment histories. In the case of single treatment per unit (the elections), this means that eligible matches must have zero treatments over the pre-treatment crossover period, as in the treated unit. In the multiple treatment scenarios (the rallies and the protests), we perform an inexact match based on the similarity of number of treatments that occur in the pre-treatment crossover period between a potential match and the treated unit. In particular, we discretized the number of treatments in the pre-treatment crossover period, and we allow a match when the potential match and the treated unit fall into the same treatment category. For the BLM protests, we defined the categories as: both have 0 prior treatments; both have 1-2 prior treatments; both have 3-9 prior treatments; and, both have 10 or more prior treatments. Separately, for the rallies, categories were defined similarly, but with additional accounting for the level of treatment exposure to neighboring counties (See Section 6.1).

### 4.2 Matching refinement via distance matching based on chosen covariates

After applying these conditions on the treatment history, we conduct distance matching based on a set of county-level time-invariant demographic characteristics, time-varying social distancing mobility data, and features of the epidemic at the county level relevant to behavior and the spread of COVID-19.

Generally, the length of the matching period for time-varying covariates is subject to a bias-variance tradeoff: choosing a longer period on which to match decreases the possible bias of the model, insofar as treated and matched counties are increasingly similar with respect to the overall COVID-19 trend. However, a longer lag period leads to greater difficulty in matching, increasing the variance of the estimator.

After selecting eligible control counties, we look for counties that are similar to the treated county with respect to demographic and epidemiological characteristics from 30 days before up to 1 day before the primary election. We pick this fixed period to ensure that we always match on a 30-day period, regardless of $F$. We apply the fixed window since the covariate histories additionally reflect general information about a county's tendency to have interpersonal contact (for the mobility covariates) and information about where the county falls with respect to its epidemic trajectory (in the case of the cumulative death rate). We do not use as the window for covariate matching the pre-treatment crossover period because this shortens as $F$ increases. Using the pre-treatment crossover window as the sole basis for covariate matching would imply that unit similarity over the fixed length period decreases in importance as $F$ increases.

In the matching procedure, we include movement specifically at full-service restaurants, grocers, and fitness and recreation facilities, because they have been identified as high-risk locations for COVID-19 transmission[43,47]. We do not explicitly include indicators for NPIs (e.g., stay-at-home-orders or there retractions) or other large-scale events. We note that there is no comprehensive data source on closures or their lifting at the county level. Moreover, these orders were haphazard and not consistently enforced. That is, a stay-at-home order in one county may have been highly effective, but ignored in another[119] (these trends likely relate to political and demographic characteristics like racial composition, SES, and political views, for which we do adjust). Furthermore, adherence to the NPIs is crucial to their role in altering transmission. Consequently, it has been argued that mobility patterns are more directly linked transmissibility than the NPIs, and may be used in place of the NPIs themselves[120] or to evaluate their effectiveness[121]. Mobility is frequently used in models of COVID-19 transmissibility[122]. In line with this, numerous studies have found a strong link between transmissibility, NPI measures, and mobility patterns[45,123–126]. Mobility is not a perfect measure of transmissibility, and has the strongest relationship to transmissibility during the earlier phases of the pandemic when NPIs were being aggressively implemented[35,127,128]. Consequently, in light of these considerations, we believe we effectively adjust (to some extent) for confounders that could arise from NPIs or other large-events, especially in tandem with our other covariates. Specifically, we focus on broad mobility categories that have been demonstrated to be linked to COVID-19 transmission: visits to restaurants, fitness and recreation facilities, and grocery stores[43] which we believe are relevant to our primary analysis of large-scale gatherings in general.

Additionally, we use the following county-level socio-demographic variables: (log) median income, percentage 65 years and above, percent African American, percent Hispanic, the population density, and Donald Trump's share of the vote in the 2016 presidential election. These demographic characteristics may be relevant to how people behave, which in turn may affect the spread of the virus: e.g., counties in which Trump had a higher share of the vote may be less likely to practice social distancing; counties with a higher percentage of African Americans or that are lower in terms of median income may have a greater share of essential workers[11,129]. Additionally, to adjust for the features of the epidemic, we adjust for the date of the first reported infection in that county as well as one time-varying feature, namely, the cumulative county-level death rate. Note that matching on the cumulative death rate implicitly contains information about the whole history of the epidemic trend, adjusting not only for the daily deaths over the lag period, but also the total deaths accrued over the reporting window. We think that this adjustment better matches counties in terms of their COVID-19 trends. Finally, we additionally match on the county-level propensity to wear a mask for the GA special election, and the NJ and VA gubernatorial elections, which took place after this survey data was collected, to further adjust attitudes and behavior regarding COVID-19.

Under matching refinement, we select at most five units that are the most similar to the treated unit during the matching period in terms of Mahalanobis distance:

$$(3)\ S_{i,t}(i') = \frac{1}{L}\sum_{l=1}^{L}\sqrt{\left(V_{i,t-l-V_{i',t-l}}\right)^T \Sigma_{i,t-l}^{-1}\left(V_{i,t-l} - V_{i',t'l}\right)}$$

where the beginning of the matching window before a treatment, is given by *L = 30.* We note that $i' \in M_{i,t}$ is a unit in the set of potential matches to a treated unit $i$, and $V_{i,t'}$ is the set of time-varying covariates for which we adjust in the pre-treatment lag period, and $\Sigma_{i,t'}$ is the diagonal of the sample covariance matrix. The distance is computed for each potential match, for each day in the pre-treatment lag period, which is then averaged over that period (from $t - 30$ up to $t - 1$).

Subsequently, we calculate the covariate balance as the standardized mean difference between the treated and control units, across all counties used in the analysis:

$$(4)\ B_{it}(j,l) = \frac{V_{i,t-l,j} - \sum_{i' \in M_{i,t}} w_{i,t}^{i'} V_{i',t-l,j}}{\sqrt{\frac{1}{N_1 - 1}\sum_{t'=1}^{N}\sum_{t'=L+1}^{T} D_{i',t'}\left(V_{i',t'-l,j} - \bar{V}_{t'-l,j}\right)^2}}$$

where $N_1$ is the total number of treated observations, and $V_{i,t,j}$ represents the $j$th covariate of unit $i$ on day $t$. The quantity $w_{i,t}^{i'}$ represents the weight given to match $i'$ for control unit $i$, at $t$. Note that $w = 0$ for each $i'$ that is not a match for the control unit. Finally, $\bar{B}(j,l)$ is simply the average balance over the treated units:

$$(5)\ \bar{B}(j,l) = \frac{1}{N_1}\sum_{i=1}^{N}\sum_{t=L+1}^{T} D_{i,t}\, B_{i,t}(j,l)$$

Where $N$ is the total number of units, $T$ is the total number of periods, $D_{i,t}$indicates whether unit $i$ is treated at $t$.

### 4.3 Caliper matching

Additionally, we impose a caliper to ensure sufficient covariate balance for estimation. We calculate the standardized Euclidean distance between each pair of matching covariates given by the treated and control units over the matching period and take the matching period average. We then apply a caliper threshold to a chosen set of covariates, rejecting a potential match unit if its distance is above the threshold on a chosen covariate. Caliper values are specified in terms standardized Euclidean distances between the values of specific matching covariates.

In particular, calipers were applied individually to specific matching covariates as needed, to ensure that covariate (im)balance remains below 0.1, on average, over the matching period. In addition, all models include a caliper of 0.25 on the cumulative death rate (or cumulative case rate as appropriate to the model), since this is the crucial covariate to ensure that the epidemic trajectories are similar between the treated units and their matches. The value of 0.25 is a standard recommendation in the matching literature[130–132]; however, we check sensitivity to stricter calipers in the Supplementary Information (Section 2.5). This procedure will remove potential matches since they may fail to satisfy the distance requirement. When a treated observation retains zero matches, it is removed entirely from estimation. Thus, a caliper may significantly reduce the number of treated units. For each set of model results, we present the number of treated observations that remain after the application of a caliper out of the initial number (see below for these details). In cases where we were unable to find a set of matches to the treated units with a sufficiently low balance score (below 0.1 on average over the pretreatment matching period), we dropped the mobility covariates, and matched on the remaining variables. Plots that display the balance scores for each of the models in the main text are available in the Supplementary Information (Section 3). The mobility models have the same balance scores as the respective, death rate models for each event.

This was the case for ten models: the Trump vote-share, in-person turnout, and mask behavior stratified models and the overall model for the GA election; the NJ & VA gubernatorial election model, the model for Donald Trump's rallies accounting for spillover effects; the model for Donald Trump's rallies that stratifies by Donald Trump's share of the vote in 2016 and accounts for spillover effects; the number-of-recent-protests stratified model, the protest-size stratified model, and the overall model for the protests; and the transmissibility model for the GA special election. Additionally, the date of first case was dropped from the NJ and VA election analyses since those occurred in a much later phase of the pandemic; also, visits to full-service restaurants was the only included mobility variable in this model, since we were unable to find sufficient balance with the full set of matching covariates.

## 5 Difference-in-difference estimator and Bayes Factors

For $N$ units, and $T$ total days in a dataset, our difference-in-difference estimator of the ATT is given by

$$(6)\ \hat{\delta}(\mathrm{F}) = \frac{1}{\sum_{i=1}^{N}\sum_{t=1}^{T} D_{i,t}} \sum_{i=1}^{N}\sum_{t=1}^{T} D_{i,t}\left\{\left(Y_{i,t+F} - Y_{i,t-1}\right) - \sum_{i\prime \in M_{i,t}} w_{i,t}^{i\prime}\left(Y_{i\prime,t+F} - Y_{i\prime,t-1}\right)\right\}$$

which captures the average of the average difference in the change in the daily death rate ($Y_{i,t}$), from the day before treatment ($t$ - 1) to a post-treatment time point ($t$+$F$) between a treated unit and its control counties. $D_{i,t} = 1$ defines a unit, $i$, treated at $t$ with at least one matched control unit. $i' \in M_{i,t}$. $w_{i,t}^{i\prime} = \frac{1}{|M_{i,t}|}$ is the match weight, with the number of matches in the set of matches to the treated observation denoted by $|M_{i,t}|$. We construct confidence intervals with a weighted block-bootstrap procedure appropriate for panel data with a fixed number of matches, which accounts for the number of times that a unit is used as a match and samples units rather than observations[133] (details presented below). All confidence levels presented are at the 95% significance level.

The overall estimates presented for each of the main models are simple arithmetic means of the individual estimates for each day in the post-treatment outcome window. The confidence intervals for the overall estimates are taken from aggregating the bootstrap distribution across the individual estimates (see Section 8 below). Additionally, for key models, we present Bayes Factors to provide more information about the relative evidence of the null and alternative hypotheses in our analyses. In contrast to p-values, the Bayes Factor is able to quantify evidence in favor of the null hypothesis[134]. In each case, we calculate the Bayes factor for the alternative hypothesis ($B_{10}$), where increasingly small numbers below 1 indicate stronger evidence for the null hypotheses over the alternative. We calculate the Bayes Factors using Gaussian quadrature, using the estimated mean and variance from the bootstrap distribution for the ATT (see below) to calculate a statistic for a classical t-test[135,136].

## 6 Spillover effects

Voting is a fundamentally *local* activity, and individuals typically do not travel outside of their neighborhood to cast a ballot, let alone across county lines. Furthermore, mobility in the spring and summer of 2020 was below usual levels, in the wake of stay-at-home orders and an increase in the number of individuals who work from home. With respect to the BLM protests, we note that larger protests are more likely to occur in highly urban areas, with a larger concentration of young individuals whose politics tend to align with the protests themselves. In contrast, surrounding suburban areas tend to be more Republican leaning (Supplementary Fig. 36), and less likely to have high concentrations of protesters. Thus, we assume that the protests are also an activity that is local in nature. Consequently, we believe that it is unlikely for an election or protest held in a single county to yield an increase in infections in other counties, over the study follow-up period we consider.

While assumption (1) (above in Section 3) is clearly not justified in general for the spread of COVID-19, we assume that it holds for the election and protest events over the time horizon we use. Specifically, we assume that it does not hold in the case of Trump rallies, as individuals may have traveled from surrounding counties to attend the rallies. This is especially plausible given that the rallies themselves tend to be in more Democratic-leaning areas than surrounding counties (Supplementary Fig. 36). We allow for the presence of spillover effects up to three counties away from the location of a Trump rally, on the logic that individuals may travel up to several hours to attend a rally.

We generalize the ATT to account for interference from neighboring counties (Supplementary Fig. 37). We denote by $G_{i,t}$ the exposure variable representing whether unit *i* at time *t* is directly exposed to treatment or indirectly exposed to the treatment of neighboring counties. Specifically, when $G_{i,t} = 1$, a unit *i* is directly treated at *t*, having an event within its county borders at time t, regardless of the treatment of neighboring counties; when $G_{i,t} = 2$, county *i* is not directly treated but it is the first degree neighbor of a county that holds an event at *t* (geographically adjacent); when $G_{i,t} = 3$, county *i* is not directly treated and does not have adjacent counties treated, but it is the second degree neighbor of a county that holds an event at $t$ (one county away); when $G_{i,t} = 4$, county $i$ is the third degree neighbor of a county that holds an event at $t$ (two counties away). $G_{i,t} = 0$ when a county $i$ at day $t$ is not directly treated and is at least three counties away from one that holds an event at $t$.

Consequently, we take $G_{i,t}$ the treatment exposure for unit $i$ *on day* $t$, valued $G_{i,t} > 0$ when the unit is treated on day $t$ (it holds or is near an event of interest), and 0 otherwise. Based on this outcome window, we assume that the potential outcome of each unit $i$ at day $\tau$ depends on the event history in a carry-over window, that is, $F_{max}$ days to $F_{min}$ days before $T$, with $F_{max} > F_{min}$. That is, the outcome at $\tau$ depends on interventions $G_{i,q}$, with $q \in \{\tau - F_{max}, \dots 0, \dots, \tau - F_{min}\}$ i.e., $Y_{i,\tau}\left(\{G_{i,q}\}_{q=\tau-F_{max}}^{\tau-F_{min}}\right)$.

In this setting, the ATT becomes:

$$(7)\ \delta(g, F;\ F_{min}, F_{max}) = E\left[Y_{i,t+F}\left(G_{i,t} = g, \{G_{i,t-l_1}\}_{l_1=1}^{F_{max}-F}, \{G_{i,t+l_2}\}_{l_2=1}^{F-F_{min}}\right) - Y_{i,t+F}\left(G_{i,t} = 0, \{G_{i,t-l_1}\}_{l_1=1}^{F_{max}-F}, \{G_{i,t+l_2}\}_{l_2=1}^{F-F_{min}} = 0\right) \mid G_{i,t} = g\right]$$

where $g \in \{1,2,3,4\}$, represent the treatment exposure category. Here, 1 represents direct exposure to treatment; values from 2 to 4 represent indirect exposure, via neighboring treatments.

While estimation of the total effect includes spillover from treated units onto treated units[137], we ignore this possibility, simply averaging within different exposures among treated units. This assumption is plausible if we assume that many attendees travel from Republican-leaning counties to the rally locations, and that attendees do not typically attend more than one rally on a single day. After matching, four sets of ATTs are estimated, for each level of exposure.

### 6.1 Selection of eligible matches based on treatment history with spillover effects

As above, (Section 4.1), we discretized the number of treatments in the pre-treatment crossover period, and we allow a match when the potential match and the treated unit fall into the same treatment category. For the rallies, categories were defined similarly, but with additional accounting for the level of treatment exposure to neighboring counties. For treated units, eligible matches were selected based on treatment history criteria that accounted for exposure. Specifically, we allow a match if it is in the same category, for each exposure type, as the treatment, where the categories are defined has having zero vs. one or more prior treatments within 30 days of the treatment.

## 7 Stratification

We additionally stratify our estimation based on properties of the treated observations (e.g., population density, in-person turnout rate). These models apply the standard matching procedure, but separate treated observations and their matches into different strata; the ATTs in each stratum are estimated separately.

## 8 Bootstrap procedure

The entire time series of a unit is resampled, to account for within-unit dependence over time, in a weighted block-bootstrap design. This method accounts for matching with a fixed number of units, and the use of a matched unit for more than one treated observation, through the calculation of observation weights:[37]

$$(8)\ W_{i,t}^{*} = \sum_{i'=1}^{N} \sum_{t'=1}^{T} D_{i',t'}\, v_{i,t}^{i',t'}$$

where

$$v_{i,t}^{i',t'} = \begin{cases} 1 \text{ if } (i,t) = (i', t' + F) \\ -1 \text{ if } (i,t) = (i', t' - 1) \\ -w_{i,t'}^{i} \text{ if } i \in M_{i',t'} \ \& \ (i', t' + F) \\ w_{i,t'}^{i} \text{ if } i \in M_{i',t'} \ \& \ (i', t' - 1) \\ 0, \quad \text{otherwise} \end{cases}$$

an observation's weight is assigned to 1 if it is a treated observation and it falls within the outcome window. It is valued at -1 in the case of the period immediately prior to the focal treatment. An observation that is not treated is assigned a non-zero weight $w_{i,t}^{i'} = \frac{1}{|M_{i,t}^{*}|}$ when $i' \in M_{i,t}^{*}$, i.e., when they fall into the outcome window, or the period preceding treatment, for a focal treatment observation. Imai et al. (2021)[37] note that the estimator may alternatively be calculated using the observation weights, as:

$$(9)\ \frac{\sum_{i=1}^{N} \sum_{t=1}^{T} W_{i,t}^{*}\, Y_{i,t}}{\sum_{i=1}^{N} \sum_{t=1}^{T} D_{i,t}}$$

consequently, in the bootstrap, we sample units and then calculate this quantity for each bootstrap sample. We take our confidence intervals from percentiles 2.5 and 97.5 of the distribution formed from 10000 resamples.

## Supplementary Information

Supplementary Information is enclosed as a separate PDF.

## Data Availability

The data on COVID-19 case and death counts are available from the Johns Hopkins Coronavirus Resource Center (https://coronavirus.jhu.edu/data). US Census data is available also publicly available (https://www.census.gov/programs-surveys/acs). The mobility tracking data is available from SafeGraph, Inc. and is freely available to academic researchers (https://www.safegraph.com/products/places). The elections turnout data is available directly from state governmental election agencies. The protest event data is available as the US Crisis Monitor dataset from the Armed Conflict Location & Event Data Project (https://acleddata.com/special-projects/us-crisis-monitor/). County-level masking data is available from the New York Times GitHub repository (https://raw.githubusercontent.com/nytimes/covid-19-data/master/mask-use/mask-use-by-county.csv).

## Code Availability

Data cleaning and preparation was carried out with the R programming language[110]. All analysis was conducted in the Julia programming language[109] with the "TSCSMethods" Julia package (https://github.com/human-nature-lab/TSCSMethods.jl), additionally paper replication materials are available on GitHub as the "COVIDPoliticalEvents" Julia package (https://github.com/human-nature-lab/COVIDPoliticalEvents.jl).

## Acknowledgements:

We thank Ben Snyder for help with data development. We thank Gary King for helpful comments. This research was supported by the Robert Wood Johnson Foundation. The funders had no role in study design, data collection and analysis, decision to publish or preparation of the manuscript.

## Author Contributions:

Conceptualization: EF, LF, MA, NAC
Methodology: EF, LF, MA, NAC
Statistical Analysis: EF, LF
Funding acquisition: NAC
Writing: EF, LF, MA, NAC

## Competing Interests:

Authors declare that they have no competing interests.

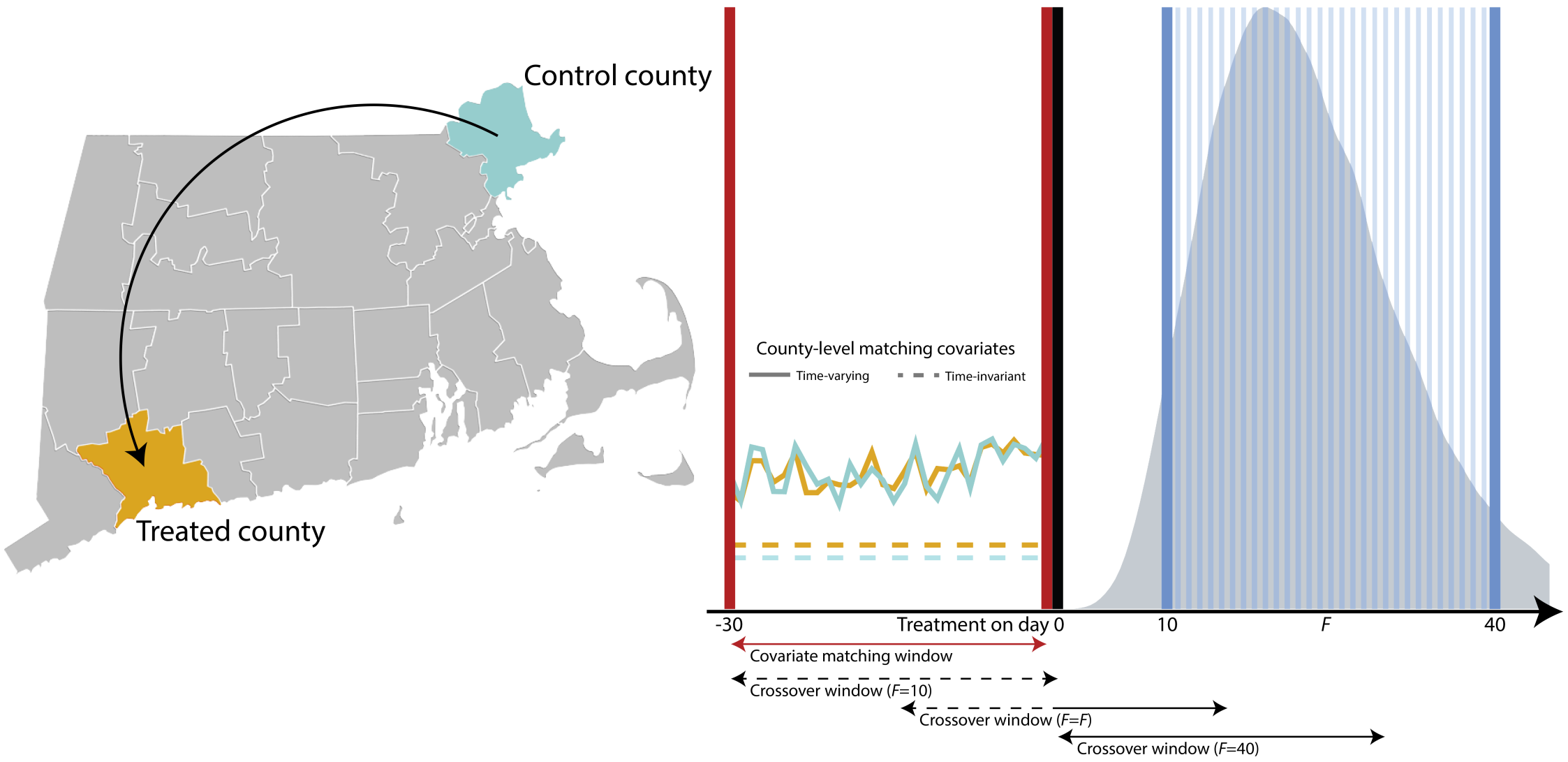

Figure 1: Overview of matching and estimation. We estimate the "average treatment effect on the treated" (ATT) for each day in the outcome window, from 10 to 40 days after a treatment event (an election, rally, or protest), which corresponds to the 5th and 95th percentiles of the empirically observed distribution of times from infection to death for SARS-CoV-2 (vertical blue lines). Given this, we use the same time window for the primary analyses of the other outcomes. Treated counties are matched to, at most, the five best control counties based on the similarity of their covariate values from 30 days to 1 day before a treatment event, i.e., the "covariate matching window" (red solid line arrow) – although here we show just one matched county for illustration. Treated counties may have between one and five matches, contingent on the quality of available matches. Covariates are either time-varying (e.g., the cumulative death rate, mobility based on phone data) or static (e.g., population density) over our study horizon. The "crossover window" is the period in which we examine the treatment history of potential matched control counties. A control county may not be treated during the post-treatment portion of the crossover window (black solid line arrows), defined as 30 days before F, up to the day before treatment. Eligible control units must have treatment histories that are similar to the treated unit in the pre-treatment crossover window, 30 days before F to 10 days before F (when defined for the post-treatment period, shown by the dashed blackline arrows). That is, we do not just match on a fixed period 30 days before treatment (which is what we do for the matching covariates), but we also deal with the treatment histories in a nuanced manner. Finally, calipers were applied individually to specific matching covariates as needed, to ensure that covariate (im)balance remains below 0.1, on average, over the matching period. Treated observations were dropped from the analysis if they had no good-quality matches.

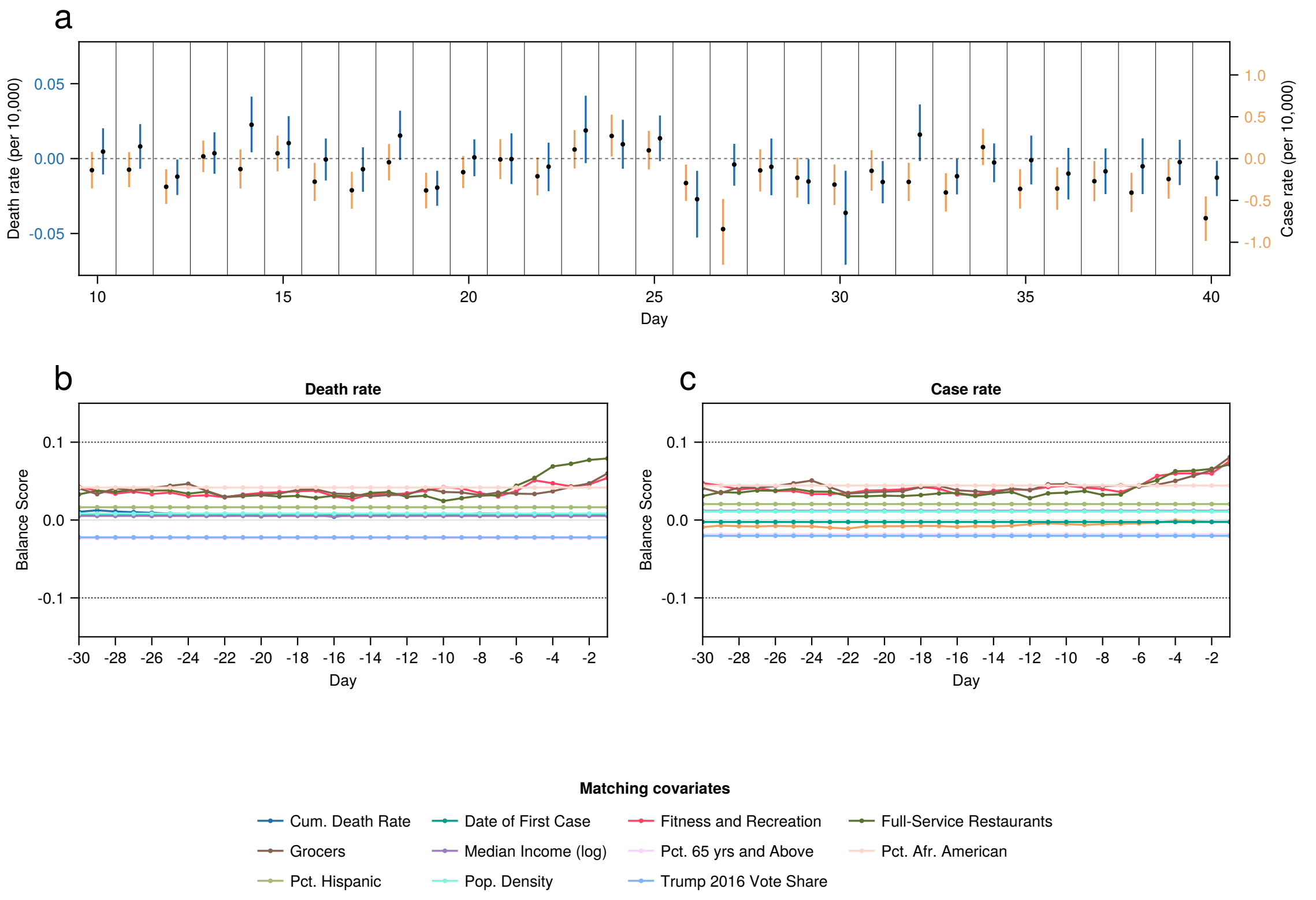

Figure 2: Impact of all mass gatherings for political expression on COVID-19 mortality and case rates in the USA. Generally, we observed no statistically significant increase, on average, for treated counties from a period of 10 to 40 days after a primary election was held for either outcome. (A) Overall "average treatment effect on the treated" (ATT) estimates for the political events, representing the average difference in the change in death (blue) and case (orange) rates, from the day before treatment to 10 to 40 days after an election. The error bars indicate the 95% CIs. For the death rate outcome, for each day on 10 to 40, N = 1493, 1486, 1477, 1474, 1472, 1467, 1461, 1458, 1456, 1456, 1455, 1451, 1450, 1449, 1445, 1445, 1444, 1443, 1442, 1442, 1442, 1434, 1433, 1433, 1433, 1433, 1433, 1433, 1431, 1430, and 1429. N is the number of treated counties on a day in the outcome window. There are an average of 6085 matched units across the period. For the death rate outcome, for each day on 10 to 40, N = 1680, 1672, 1663, 1661, 1656, 1652, 1649, 1646, 1645, 1644, 1642, 1640, 1638, 1637, 1636, 1636, 1634, 1633, 1632, 1631, 1631, 1621, 1621, 1620, 1620,1619, 1619, 1619, 1618, 1618, and 1618. N is the number of treated counties on a day in the outcome window. There are an average of 7277 matched units across the period. (B) Covariate balance for the death rate model. The balance scores reflect the similarity between the treated units

and their matched counties for 30 days before up to 1 day before a political event. They reflect the covariate balance after matching refinement and the application of a caliper to ensure match quality. The balance scores are, on average over the matching window, within the threshold of 0.1, indicating sufficient similarity between the treated and matched counties for the estimated ATTs. (C) Covariate balance for the case rate model. The balance scores are, on average over the matching window, again within the threshold of 0.1. See Extended Data Figs. 2-3 and Methods for more details.

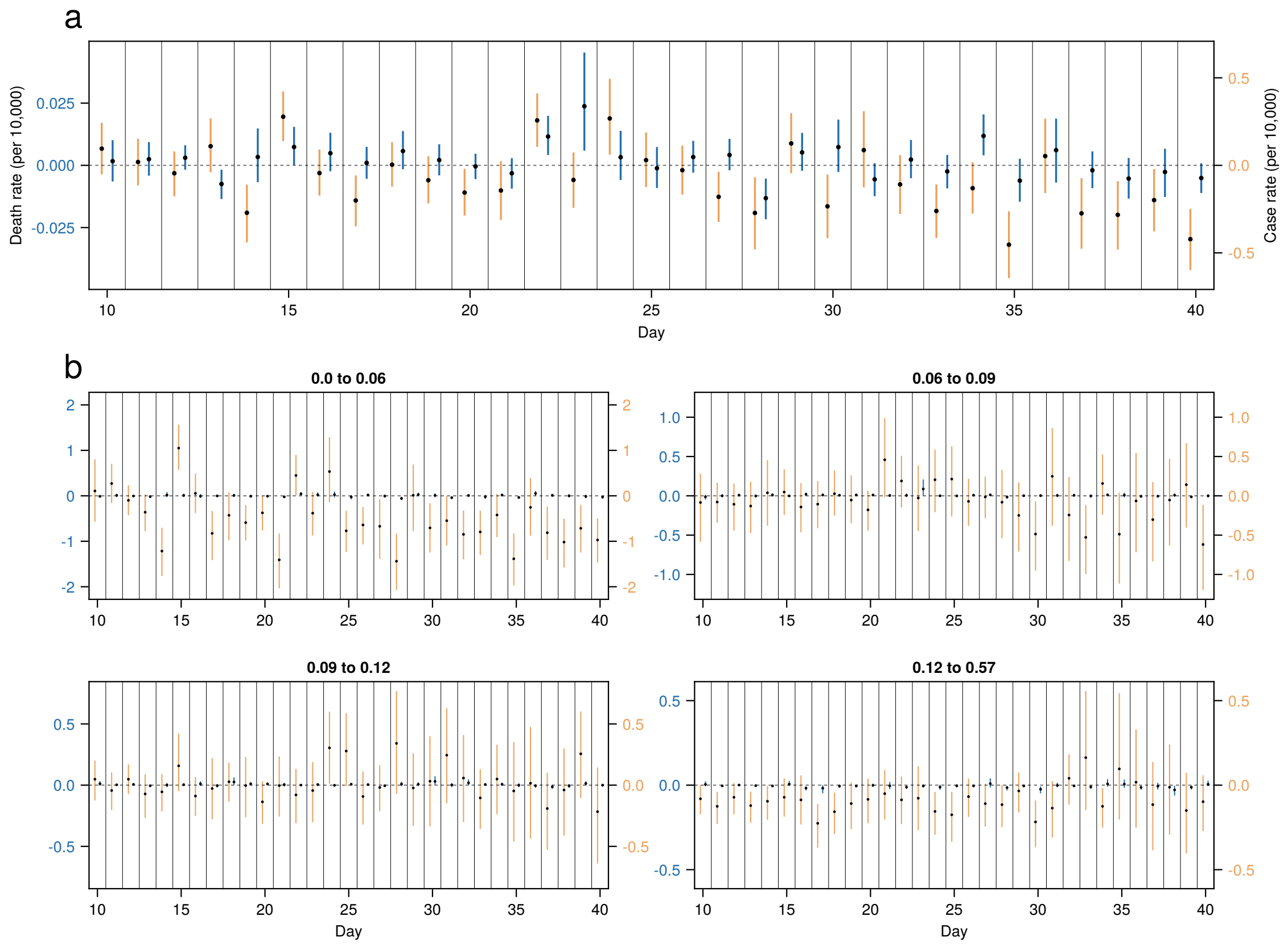

Figure 3: Impact of the primary elections on COVID-19 mortality and case rates. Generally, we observed no statistically significant increase, on average, for treated counties from a period of 10 to 40 days after a primary election was held. In each panel, the error bars indicate the 95% CIs. (A) Overall “average treatment effect on the treated” (ATT) estimates for the primary elections, representing the average difference in the change in death (blue) and case (orange) rates, from the day before treatment to 10 to 40 days after an election. For the death rate outcome, N = 965 for days 10-23, 961 for 34-30, 956 for 31 to 40. For the case rate outcome, N = 1057, 1056, and 1051 for the same ranges in the outcome window. N represents the number of treated units. The average number of matched units are 4284 and 4878 for the death and case rates, over the whole period. (B) ATT estimates, stratified by the in-person voter turnout rate (defined as the number of in-person voters out of the total county population). The strata represent quantiles of the distribution of turnout rates over the counties with elections. Since turnout data were not available

for CO, IA, KY, MA, MI, MN, MS, MO, MT, NV, NY, OK, PA, RI, SD, UT, WV, these states were estimated as a separate category (see Methods). For the death rate, for each turnout bin N = 132, 141, 121, 88 on days 10-23; 131, 141, 121, 88 on days 24-30; 130, 140, 121, 88 on days 31-40. N represents the number of treated counties on a day. For the case rate, N = 567, 664, 646, 426 for days 10-16; 566, 664, 646, 426 for days 17-23; 566, 662, 646, 426 for days 24-30; 557, 656, 640, 424 for days 31-40. The average number of matched units over the period are 470, 563, 548, 416 for the death rate, and 154, 156, 142, 90 for the case rate in each stratum. Covariate balance for the results in (A) and (B) are, on average over the matching window, within the threshold of 0.1, indicating sufficient similarity between the treated and matched counties for the estimated ATTs (See Extended Data Fig. 4; Supplementary Figs. 38-42).

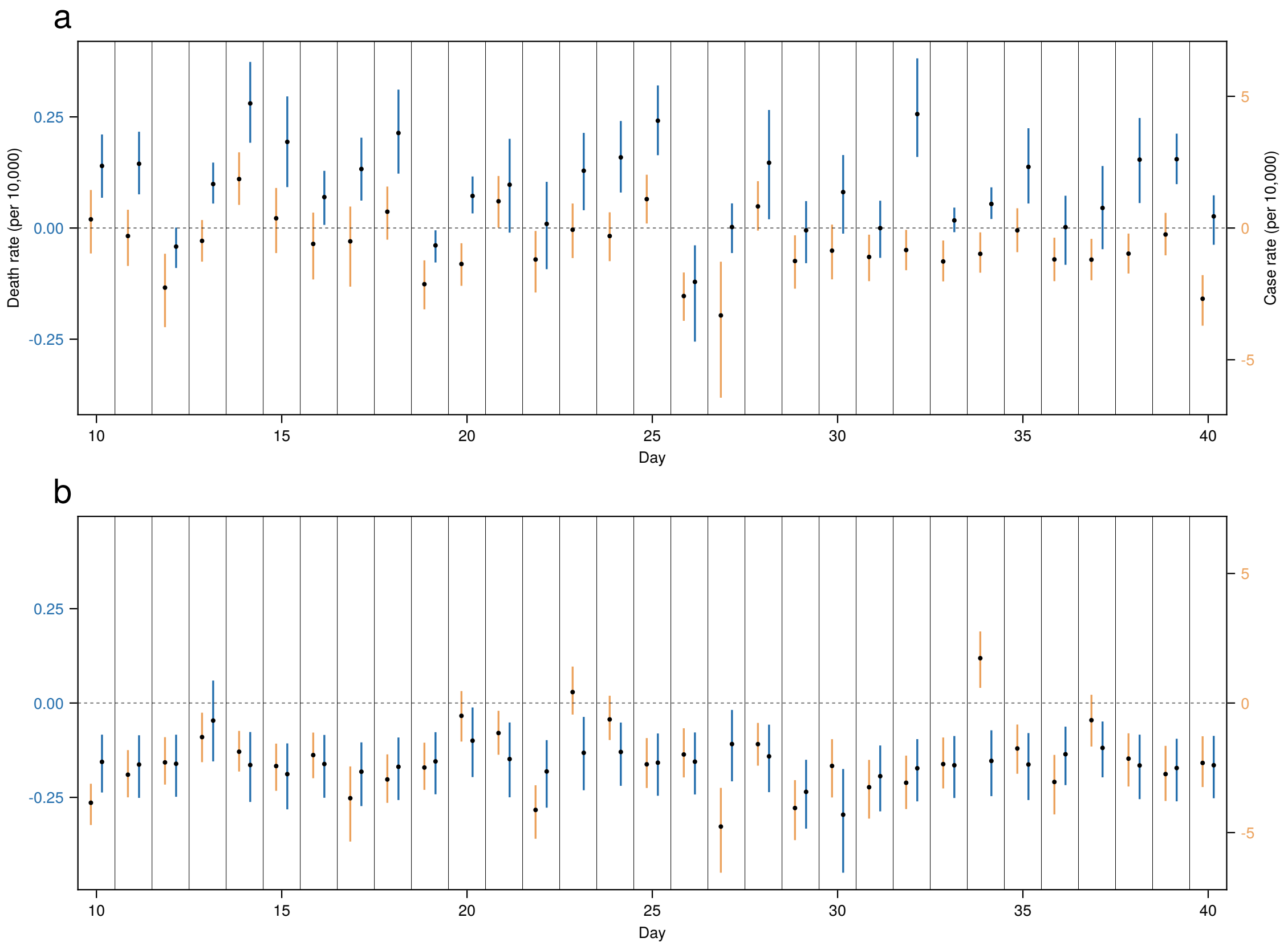

Figure 4: Impact of the Georgia special election and the New Jersey and Virginia gubernatorial elections combined on COVID-19 mortality and case rates. In both panels, the error bars indicate the 95% CIs. (A) Overall "average treatment effect on the treated" (ATT) estimates for the GA election, representing the average difference in the change in death (blue) and case (orange) rates, from the day before treatment to 10 to 40 days after the election. While there is some evidence for a very modest increase in risk of local mortality in the follow-up window, we do not find a statistically significant average effect over this period (Table S2); furthermore, we find no evidence for a similar general increase in the case rates. For the death rate, N = 137 for each day in the outcome window, where N represents the number of treated counties on a day in the outcome window. For the case rate, N = 152. Respectively, there are 566 and 701 matched counties. (B) Overall ATT estimates for the two gubernatorial elections (in NJ and VA), representing the average difference in the change in death rates, from the day before treatment to 10 to 40 days after the election. There is some evidence for a very modest decline in risk of local mortality and in the case rates and case prevalence on most days in the follow-up window, but we do not find a statistically significant overall average effect over this whole period (Supplementary Table 2). For the death rate, N = 125 for each day in the outcome window, where N represents the number of treated counties on a day in the outcome window. For the case rate, N = 141. Respectively, there are 528 and 649 matched counties. Covariate balance for the results in (A) and (B) are, on average over the matching window, within the threshold of 0.1, indicating sufficient similarity between the treated and matched counties for the estimated ATTs (See Extended Data Fig. 6; Supplementary Figs. 43-46).

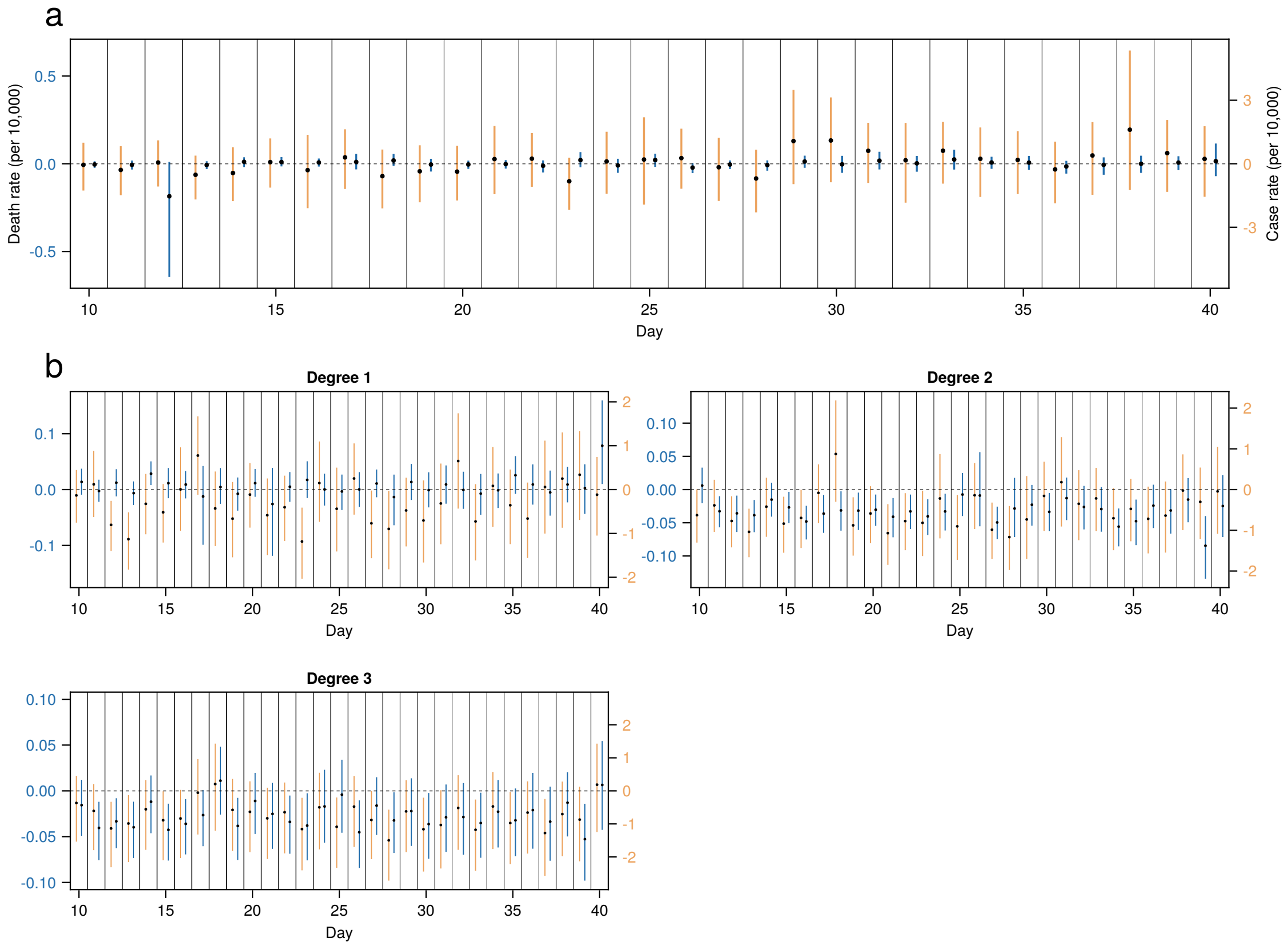

Figure 5: Impact of Donald Trump's rallies on COVID-19 mortality and case rates. Generally, we observed no statistically significant increase, on average, for treated counties from a period of 10 to 40 days after a political rally was held. In each panel, the error bars indicate the 95% CIs. (A) Overall "average treatment effect on the treated" (ATT) estimates for Donald Trump's rallies, representing the average difference in the change in death (blue) and case (orange) rates, from the day before treatment to 10 to 40 days after a rally. For the death rate, N = 60 for day 10, 59 for 11-13, 58 for 14-40, where N is the number of treated counties. The average number

of matched units are, respectively, 252 and 303. (B) ATT estimates, for counties, stratified by indirect exposure to the treatment. Degree 1 represents counties adjacent to a county that held a rally, degree 2 gives a county 2 county away from one that held a rally, and degree 3 represents a county two counties away from one that held a rally. For the death rate for degrees 1, 2, and 3, N = 366, 720, 1069 for day 10; 365, 719, 1064 for days 11-14; 364, 718, 1060 for day 15; 363, 718, 1058 for day 16; 363, 718, 1056 for day 17; 363, 715, 1056 for day 18-19; 363, 715, 1054 for day 20; 363, 714, 1053 for day 21-22; 363, 713, 1053 for day 23-25; 363, 711, 1053 for day 26-30; 363, 710, 1052 for day 31-38; 363, 709, 1049 for day 39; and 363, 708, 1047 for day 40. N represents the number of treated counties. Over the whole period, the average number of matched counties is 1679, 3291, 4820. For the case rate for degrees 1, 2, and 3, N = 384, 760, 1110 for day 10; 383, 760, 1105 for day 11-12; 383, 759, 1104 for day 13; 382, 759, 1104 for day 14; 382, 758, 1104 for day 15-16; 381, 758, 1102 for day 17-18; 381, 757, 1102 for day 19-20; 381, 757, 1101 for day 21-22; 380, 757, 1101 for day 23-29; 380, 756, 1101 for day 30; 380, 755, 1101 for day 31-35; 380, 753, 1101 for day 36-38; 380, 751, 1100 for day 39; and 380, 750, 1098 for day 40. Over the whole period, the average number of matched counties is 1840, 3634, 5262. Covariate balance for the results in (A) and (B) are, on average over the matching window, within the threshold of 0.1, indicating sufficient similarity between the treated and matched counties for the estimated ATTs (See Extended Data Figs. 7-8; Supplementary Figs. 47-48).

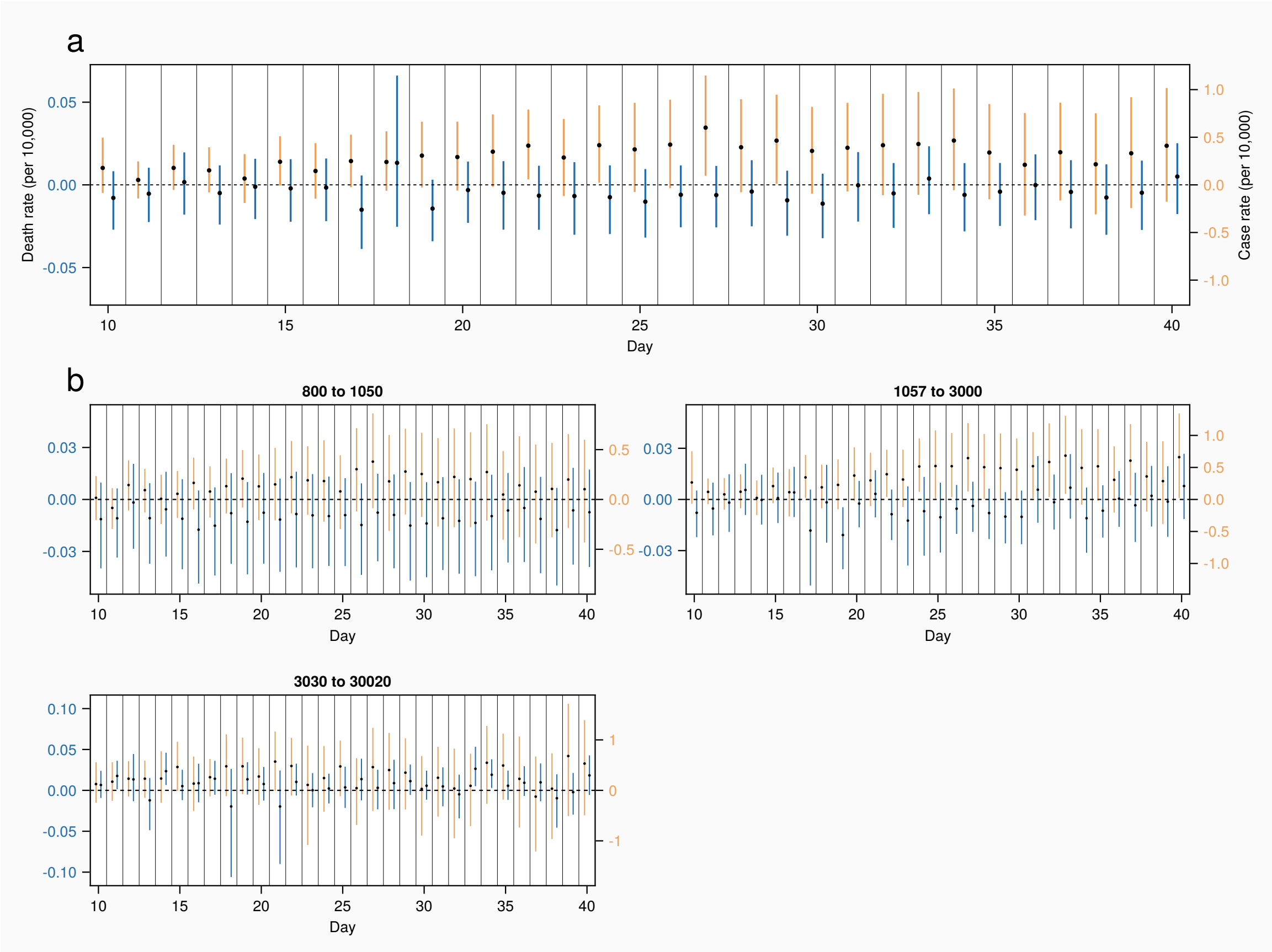

Figure 6: Impact of the BLM protests on COVID-19 mortality. Generally, we observed

no statistically significant increase for treated counties from a period of 10 to 40 days after a protest was held. In each panel, the error bars indicate the 95% CIs. (A) Overall "average treatment effect on the treated" (ATT) estimates for the protests, representing the average difference in the change in death (blue) and case (orange) rates, from the day before treatment to 10 to 40 days after a protest. For the death rate, N = 402, 385, 377, 369, 361, 357, 354, 352, 350 for days 10-18, 349 for day 19-24, 348 for day 25, 347 for day 26-29, 346 for day 30, 345 for day 31, 344 for day 32-34, 343 for day 35-37, 342 for day 38-40. N represents the number of treated counties. The average number of matched counties over the period is 1346. For the death rate, N = 448, 441, 437, 432, 430, 424, 422, 420 for day 10-17, 419 for day 18-21, 418 for day 22-24, 417 for day 25-37, 416 for day 38-40. The average number of matched counties is 1764. (B) ATT estimates, stratified by the protest size (defined as the total crowd size estimate within the treated county on that day). The strata represent quantiles of the distribution of estimated crowd sizes. For the death rate for each protest size bin, N = 150, 181, 48 for day 10; 145, 172, 44 for day 11; 142, 168, 43 for day 12; 139, 167, 41 for day 13; 137, 164, 39 for day 14; 134, 161, 39 for day 15; 132, 160, 39 for day 16; 132, 159, 38 for day 17; 132, 157, 38 for day 18; 132, 156, 38 for day 19-23; 131, 156, 38 for day 24; 131, 156, 37 for day 25; 131, 155, 37 for day 26-29; 131, 154, 37 for day 30; 130, 154, 37 for day 31-32; 129, 154, 37 for day 33-34; 128, 154, 37 for day 35-37; 128, 153, 37 for day 38-40. N represents the number of treated counties. The average number of matched counties is 533, 545, and 130. For the case rate, N = 170, 210, 66 for day 10; 170, 204, 64 for day 11; 170, 201, 62 for day 12; 168, 199, 61 for day 13; 168, 198, 60 for day 14; 165, 196, 60 for day 15; 165, 195, 60 for day 16; 164, 194, 60 for day 17; 164, 194, 59 for day 18-21; 164, 194, 58 for day 22-24; 164, 194, 57 for day 25-37; 164, 193, 57 for day 38-40. The average number of matched counties is 699, 778, and 220 for each stratum. Covariate balance for the results in (A) and (B) are, on average over the matching window, within the threshold of 0.1, indicating sufficient similarity between the treated and matched counties for the estimated ATTs (See Extended Data Fig. 9; Supplementary Figs. 49-53).

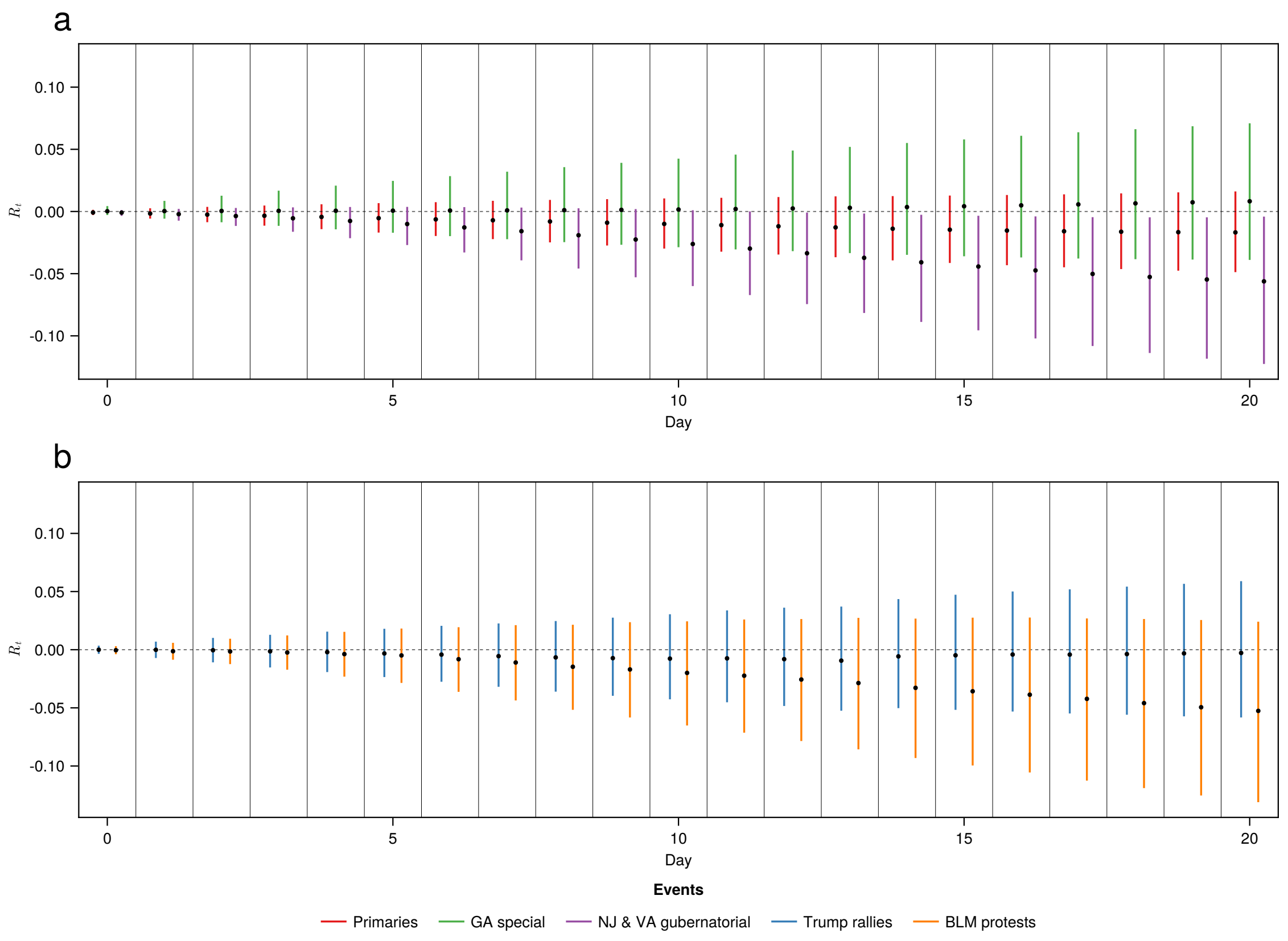

Figure 7: Impact of large-scale political events on COVID-19 transmissibility ($R_t$), representing the average difference in the change in the transmissibility of the virus, from the day before treatment to the day of a protest, up to 20 days after a protest. We find no evidence of a significant increase in transmissibility stemming from these events within 20 days of an event, starting on the day of the event (day 0). In both panels, the error bars indicate the 95% CIs. (A) Overall "average treatment effect on the treated" (ATT) estimates for the primary, GA special, and NJ & VA gubernatorial elections. (B) Overall ATT estimates for the BLM protests and Donald Trump's political rallies. For the primaries, N = 971 for day 0-13, and 969 for 14-20. N represents the number of treated counties. The number of matched counties is 4473 and 4466 for the same periods; for the GA special, N = 87, with 211 matched counties. For the gubernatorial elections, N = 83 with 230 matched counties. For the Trump rallies, N = 65, with an average of 308 matched counties. For the BLM protests, N = 466 for day 0, 460 for day 1, 458 for day 2, 456 for day 3, 543 for day 4, 452 for day 5-6, 451 for day 7, 449 for day 8-10, 448 for day 11, 447 for day 12-14, 446 for day 15-20. The average number of matched units is 1902. Covariate balance for the results in (A) and (B) are, on average over the matching window, within the threshold of 0.1, indicating sufficient similarity between the treated and matched counties for the estimated ATTs (See Supplementary Figs. 54-59).

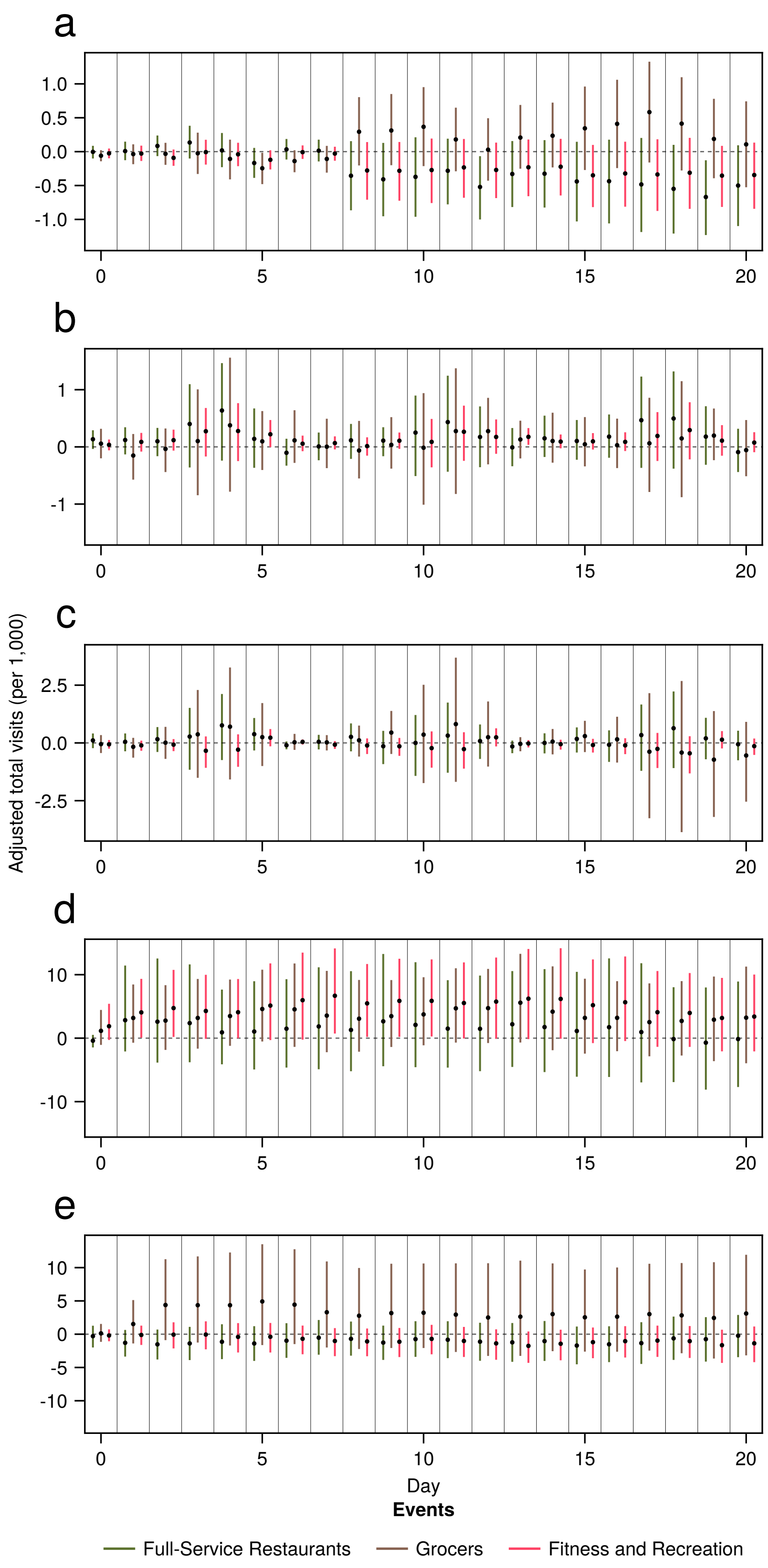
a
b
c
d
e
1.0
0.5
0.0
-0.5
-1.0
1
0
-1
2.5
0.0
-2.5
10
0
-10
10
5
0
-5
-10
0
5
10
15
20
Adjusted total visits (per 1,000)
Day
Events
Full-Service Restaurants
Grocers
Fitness and Recreation

Figure 8: Impact of large-scale political events on COVID-19 mobility, representing the average difference in the change in the number of adjusted visits to a given location type, from the day before treatment to the day of an event, up to 20 days after an event. For each event, we examine visits to full-service restaurants, grocers, and fitness and recreation centers. We find no evidence of a significant change increase in mobility stemming from these events within 20 days of an event. In each panel, the error bars indicate the 95% CIs. The panels present the overall "average treatment effect on the treated" (ATT) estimates for the (A) primaries, (B) GA special election, (C) NJ & VA gubernatorial elections, (D) Donald Trump's political rallies, and (E) the BLM protests. For the primaries, the number of treated units, N = 971 for day 0-13, and 967 for 14-20, 960 for 21-31, 957 for 32-34, 953 for 35-38, and 952 for 39-40. There is an average of 4271 matched units over the whole period. For the GA special, N = 83 for each day, with 252 matched units. For the gubernatorial elections, N = 83 with 284 matched counties. For the Trump rallies, N = 32 for days 0-4, 31 for days 5-16, 30 for days 17-31, and 29 for days 32-40. Over the whole period, there is an average of 284 matched units. For the BLM protests, N = 188, 181, 177, 176, 173, 171, 167, 164 for days 0-7; 163 for days 8-10, 162 for days 11-26, 161 for day 27, 160 for day 28, 159 for days 29-34, 157 for days 35-36, and 156 for 37-40. The average number of matched counties is 519 over the whole period. Covariate balance for the results in (A) and (B) are, on average over the matching window, within the threshold of 0.1, indicating sufficient similarity between the treated and matched counties for the estimated ATTs. The balance scores are identical to those for the corresponding models for the death rates.